\documentclass[11pt,epsf,letterpaper]{article}%
\usepackage{setspace}
\usepackage{color}
\usepackage{amsmath}
\usepackage{amsfonts}
\usepackage{verbatim}
\usepackage{amssymb}
\usepackage[dvips]{graphicx}
\usepackage{epsf}
\usepackage{subfigure}%
\providecommand{\U}[1]{\protect\rule{.1in}{.1in}}
\newcommand{\be}{\begin{equation}}
\newcommand{\ee}{\end{equation}}
\newcommand{\ba}{\begin{eqnarray}}
\newcommand{\ea}{\end{eqnarray}}

\begin{document}

\title{SU(N) Multi-Skyrmions at Finite Volume}
\author{Fabrizio Canfora$^{1, *}$, Marco Di Mauro$^{2}$, Maxim A. Kurkov$^{3}$, Adele
Naddeo$^{2}$\\$^{1}$\textit{Centro de Estudios Cient\'{\i}ficos (CECS), Casilla 1469,
Valdivia, Chile.}\\$^{2}$\textit{Dipartimento di Fisica ``E.R. Caianiello", Universit\'a di
Salerno,} \\\textit{Via Giovanni Paolo II, 84084 Fisciano (SA), Italy.}\\$^{3}$\textit{Dipartimento di Matematica e Applicazioni ``R. Caccioppoli",
Universit\'a di Napoli Federico II,}\\\textit{ Via Cinthia, 80126 Napoli, Italy.}\\$^{*}${\small canfora@cecs.cl }}
\maketitle

\begin{abstract}
We study multi-soliton solutions of the four-dimensional SU(N) Skyrme model by combining the
hedgehog ansatz for SU(N) based on the harmonic maps of $S^{2}$ into
$CP^{N-1}$ and a geometrical trick which allows to analyze explicitly
finite-volume effects without breaking the relevant symmetries of the ansatz.
The geometric set-up allows to introduce a parameter which is related to the
't Hooft coupling of a suitable large $N$ limit, in which $N\rightarrow\infty$ and the curvature of the background metric approaches zero, in such a way that their product is constant. The relevance of such a
parameter to the physics of the system is pointed out. In particular, we discuss how the discrete symmetries of the configurations depend on it.

\end{abstract}

\section{Introduction}

One of the most intriguing theoretical results in Quantum Field Theory (QFT
henceforth) has been the realization that fermions can emerge out of purely
bosonic Lagrangians as solitonic excitations (for a detailed review see
\cite{spin}). The clearest demonstration that the importance of this result
goes far beyond pure theoretical physics is given by the Skyrme theory
\cite{skyrme} which is one of the most important models of nuclear and
particle physics. The Skyrme term \cite{skyrme} allows the existence of static
soliton solutions with finite energy, called \textit{Skyrmions} (see~
\cite{multis2, manton, susy}) describing fermionic degrees of freedom (see
\cite{bala2,bala0,bala1,ANW,guada,moduli,skyrmonopole} and references
therein). The wide range of applications of the theory in other areas (such as
astrophysics, Bose-Einstein condensates, nematic liquids, multi-ferric
materials, chiral magnets and condensed matter physics in general
\cite{astroSkyrme,useful1,useful2,useful3,useful4,useful5,useful6,useful7.1,useful7,useful8}%
) is well recognized by now. It is also worth to emphasize that the Skyrme
model appears in a very natural way in the context of the AdS/CFT
correspondence \cite{sakai}.

From the point of view of nuclear physics, it is very important to have
analytic tools allowing to analyse the Skyrme model when interacting
Skyrmions are present within bounded regions. This case is relevant whenever
one wants to take into account the effects of the finite size on the
topological properties of the Skyrmions themselves. It is commonly believed that many-nucleons systems (such as the ones
occurring in nuclear pasta \cite{pasta,pasta0}; for a review see
\cite{pasta1}) are completely out of reach of the analytical techniques
provided by soliton theory already in the $SU(2)$ case (while, at a first
glance, the $SU(N)$ case is even worse). In particular, the task to compute
physical parameters of such multi-nucleon systems with analytic
multi-Skyrmionic configurations is believed to be completely hopeless.

Recently, the generalized hedgehog ansatz in the $SU(2)$ case, introduced in
\cite{canfora,canfora2,canfora3}, allowed the construction of the first
multi-Skyrmions at finite volume: namely, the first exact solutions of the
Skyrme model representing interacting elementary Skyrmions with a non-trivial
winding number, in which finite-volume effects can be explicitly taken into
account (arriving at a good prediction for the compression modulus), was obtained \cite{ccz}\footnote{Using similar techniques (see \cite{cgp} and \cite{para}), intriguing cosmological properties of the Skyrme model have been disclosed.}. The way to do this is to write the system in
a modified ``cylinder-like" metric whose curvature is parametrized by a length
$R_{0}$. Then multi-Skyrmionic configurations look like necklaces of
elementary Skyrmions interacting in bounded tube-shaped regions. The ground
state of such multi-Skyrmions has the remarkable property that, although the
BPS bound in terms of the winding cannot be saturated, a new topological
charge exists which leads to a different BPS bound, which can instead be saturated.

In this paper, we consider the $SU(N)$ case. A very powerful technique to
construct multi-Skyrmionic configurations in unbounded regions is given by an ansatz for $SU(N)$ Skyrmions introduced in \cite{su(n)1} and \cite{su(n)2}, based on harmonic maps of $S^{2}$ into $CP^{N-1}$, which allowed
the construction of many interesting numerical multi-Skyrmionic
configurations. Here we will exploit the fact that the metric we use is
spherically symmetric, just like flat space, so the harmonic map ansatz can be
used also in the present case without modifications. We shall see that in the
new metric the equations simplify with respect to those studied in
\cite{su(n)1} and \cite{su(n)2}; in particular they become autonomous, thus techniques from dynamical systems
theory become available. Thanks to the choice of the background geometry, a novel type of large $N$ limit becomes possible in which  $N$ is large and the curvature is small in such a way that their product is constant. In this way one can see that both the effects of the curvature become negligible and the field equations remain autonomous (so that, in such a limit, one can kill the curvature of the metric but keeping all the advantages of the technique of  \cite{ccz}). It is also worth to emphasize that it is precisely the large $N$ limit which discloses in the clearest possible way the role of the Skyrme model as low energy limit of QCD.
A non constructive proof of the existence of such nontrivial solutions is also provided and a few numerical solutions are exhibited as well.

This paper is organized as follows: in the second section, the $SU(2)$ case
treated in \cite{ccz} will be briefly reviewed in order to set the stage. In
the third section, the ansatz for the Skyrmions will be described. In the
fourth section, the general equations of motion are written down and the flat-large $N$ limit is
discussed. In the fifth section, we study the system  using techniques from
the theory of dynamical systems; in particular, the stability of the fixed
points in the $SU(3)$ case is analyzed, and their dependence on the geometric 't Hooft parameter is
discussed. In the sixth section, some nontrivial numerical solution of the
equations of motion in the $SU(4)$ case will be studied. In the seventh
section, some conclusions will be drawn. In the appendix, it is proved on
general grounds that the equations of motion do have nontrivial solutions
which can be interpreted as genuine $SU(N)$ Skyrmions.

\section{Generalized hedgehog at finite volume: the $SU(2)$ case}

In this section, the construction of analytic multi-Skyrmionic configurations
in the $SU(2)$ case \cite{ccz} will be shortly reviewed. The action of the
four--dimensional Skyrme model is given by
\begin{align}
& S_{Sk}    =\frac{K}{2}\int d^{4}x\sqrt{-g}\,\mathrm{Tr}\left(  - \frac{1}%
{2}R^{\mu}R_{\mu}+\frac{\lambda}{16}F_{\mu\nu}F^{\mu\nu}\right), \label{skyrmaction}\\
\ \ \ & K>0\ ,\ \ \ \lambda>0\ ,\, \ \
R_{\mu}    :=U^{-1}\nabla_{\mu}U=R_{\mu}^{i}t_{i}\ ,\ \ F_{\mu\nu}:=\left[
R_{\mu},R_{\nu}\right]  \  ,\nonumber
\end{align}
where the Planck constant and the speed of light have been set to $1$, and $K$
and $\lambda$ are the coupling constants. The $t^{i}$'s are the generators of
the flavor group, which in this section is $SU(2)$. Notice that we allowed for
curved metrics. The coupling constants $K$ and $\lambda$ are related to the
couplings $F_{\pi}$ and $e$ used in \cite{ANW}\footnote{Experimentally,
$F_{\pi}=186\ \ MeV\ ,\ e=5.45$.} by
\[
K=\frac{1}{4}F_{\pi}^{2},\quad K\lambda=\frac{1}{e^{2}}\ .
\]
The non-linear sigma model term of the Skyrme action is necessary to take into
account pions. The second term is the only covariant term leading to a
well-defined Hamiltonian formalism in time which supports the existence of Skyrmions.

The field equations following from the above action are
\begin{equation}
\nabla^{\mu}R_{\mu}+\frac{\lambda}{4}\nabla^{\mu}[R^{\nu},F_{\mu\nu}]=0.
\label{nonlinearsigma1}%
\end{equation}

The following standard parametrization of the $SU(2)$-valued scalar $U(x^{\mu
})$ will be adopted%
\begin{equation}
U^{\pm1}(x^{\mu})=Y^{0}(x^{\mu})\mathbf{1}_{2}\pm Y^{i}(x^{\mu})t_{i}%
\ ,\ \ \left(  Y^{0}\right)  ^{2}+Y^{i}Y_{i}=1\ , \label{standard1}%
\end{equation}
where $\mathbf{1}_{2}$ is the $2\times2$ identity matrix; to describe a
spherically symmetric field configuration we use the hedgehog ansatz
\cite{canfora2,ccz}:%
\begin{equation}
Y^{0}=\cos\alpha\ ,\ \ Y^{i}=\widehat{n}^{i}\sin\alpha\ ,\ \ \ \alpha
=\alpha(x,t)\ , \label{hedge1}%
\end{equation}%
\begin{equation}
\widehat{n}^{1}=\sin\theta\cos\varphi\ ,\ \ \ \widehat{n}^{2}=\sin\theta
\sin\varphi\ ,\ \ \ \widehat{n}^{3}=\cos\theta\ . \label{hedge2}%
\end{equation}

In order to mimic finite-volume effects without loosing the nice properties of
the hedgehog ansatz we will consider the following curved
background\footnote{The simplest choice of a bounded spatial metric is $S^{3}%
$, but this has already been considered in \cite{curved1,curved2}. The
geometry in Eq. (\ref{metric}) was considered in \cite{SkyrmeFirst} but with a
motivation different from the analysis of finite-volume effects. Consequently,
the main results obtained in \cite{ccz} (the derivation of both a novel BPS
bound which can be saturated and analytic multi-Skyrmions at finite volume and
an explicit formula for the compression modulus in good agreement with
experiments) are indeed novel.}
\begin{equation}
ds^{2}=-dt^{2}+dx^{2}+R_{0}^{2}(d\theta^{2}+(\sin\theta)^{2}d\varphi
^{2})\ ,\ \ -\frac{L}{2}\leq x\leq\frac{L}{2}\ , \label{metric}%
\end{equation}%
\begin{equation}
0\leq\theta\leq\pi\ ,\ 0\leq\varphi\leq2\pi\ , \label{range1}%
\end{equation}
where $L$ is the length of the $x-$interval. The total volume of space is
$V=4\pi R_{0}^{2}L$. This geometry describes three-dimensional cylinders whose
sections are $S^{2}$ spheres, so that parameter $R_{0}$ plays the role of the
(finite) diameter of the transverse sections of the tube. The fact that this
parameter replaces the radial variable $r$ in the metric also leads, as we
will see, to considerable simplification of the equations of motion, even
allowing to find exact solutions in the $SU(2)$ case. Moreover, the curvature
of this metric is proportional to $1/R_{0}^2$. As it will be explained in the next
sections, the explicit presence of this parameter in Eq. (\ref{metric}) together with the $N$ of $SU(N)$ allows
to define a smooth flat limit in which $R_{0}\rightarrow\infty\label{flatl1}$
and so all the effects of the curvature disappear (however, the global
topology\ of space remains cylindrical even in the flat limit and so it
differs from the trivial $S^{3}$ topology of flat static \textit{unbounded}
Skyrmions). Thus, in a sense, the above metric is introduced just as a ``regulator" whose local effects can be removed at the end.
In the present context, ``flat limit" really means
\begin{equation}
R_{0}\gg1\ fm\ , \label{flatl2}%
\end{equation}
so that, from the practical point of view, already when $R_{0}$ is around
$100$ $fm$ all the effects of the curvature are negligible and, consequently,
even in the flat limit in Eq. (\ref{flatl2}) finite volume effects will not
disappear. It is also worth to emphasize that the well known result that
elementary Skyrmions should be quantized as Fermions (which originally was
derived on flat spaces) has been extended to space-times with compact
orientable three-dimensional spatial sections in \cite{curvedquantization}
(and the metric in Eq. (\ref{metric}) belongs to this class).

Since, at the end, we will be interested in the flat limit in Eq.
(\ref{flatl2}), one may wonder whether it would be possible to start from the
very beginning with a flat metric. In fact, as it has been shown in
\cite{ccz}, the background metric in Eq. (\ref{metric}) is a very suitable
tool to take into account finite volume effects (since the total spatial
volume is finite) without breaking relevant symmetries of the hedgehog ansatz,
with the additional advantage of simplifying the field equations. A further
relevant advantage of the above background metric is that, unlike what happens
in the usual unbounded case, it allows to define in a very transparent way a
smooth large $N$ limit of the $SU(N)$ Skyrmions. Therefore, it is much more
convenient to analyze the Skyrme theory first within the background metric in
Eq. (\ref{metric}), and take the flat limit only later.

The effectiveness of such a choice for the metric is also shown by the results
in \cite{monopolect} and \cite{canfotalla2} in which it has been shown that, unlike what happens in
flat space, the equations for the Yang-Mills-Higgs system (in the sector with
non-vanishing non-Abelian electric and magnetic charges) possess analytic
solutions even in the case in which the Higgs coupling is non-zero.

With the above ansatz the Skyrme field equations reduce in the static case to
the following scalar differential equation for the Skyrmion profile $\alpha$
\cite{canfora2}:%

\begin{equation}
\label{SU(2)static}\left(  1+\frac{2\lambda}{R_{0}^{2}}\,\sin^{2}%
\alpha\right)  \overset{\cdot\cdot}{\alpha}- \frac{\text{sin}(2\alpha)}%
{R_{0}^{2}}\left(  1-\lambda\left[  \overset{\cdot}{\alpha}^{2}-\frac
{\text{sin}^{2}\alpha}{R_{0}^{2}}\right]  \right)  =0
\end{equation}

The winding number $W$ for such a configuration reads:%
\begin{equation}
W=-\frac{1}{24\pi^{2}}\int\epsilon^{ijk}Tr\left(  U^{-1}\partial_{i}U\right)
\left(  U^{-1}\partial_{j}U\right)  \left(  U^{-1}\partial_{k}U\right)
=-\frac{2}{\pi}\int\left(  \overset{\cdot}{\alpha}\sin^{2}\alpha\right)  dx\ .
\label{winding}%
\end{equation}
In the present case, the natural boundary conditions correspond to the choice:%
\begin{equation}
\alpha\left(  \frac{L}{2} \right)  -\alpha\left(  -\frac{L}{2}\right)
=n\pi\ ,\ \ n\in%
\mathbb{Z}
\ . \label{windinteger}%
\end{equation}
and with these boundary conditions the winding number takes the integer value
$n$. These boundary conditions are unique in that they ensure $U(-\frac{L}%
{2})=(-1)^{n}U(\frac{L}{2})$, which correspond to bosonic and fermionic states
for even and odd $n$, respectively.

Smooth solutions exist for any $n$ satisfying the above boundary conditions
for a finite range $(-L/2,L/2)$. In particular multi-soliton solutions exist,
which represent Skyrmions with winding number $n$ living in a finite spatial
volume $V=4\pi R_{0}^{2}L$. It is worth to remark that the large $n$ limit in
the present context is quite natural in order to consider thermodynamical
properties of the multi-Skyrmions system and $n$ is the baryon number:
obviously, a thermodynamical analysis only makes sense in the cases in which
the number of particles is very large.

\section{The hedgehog ansatz for SU(N)}

Now we switch to the $SU(N)$ case with generic $N$. Our analysis will be based
on the techniques introduced in \cite{su(n)1,su(n)2}, which we shall now
briefly describe. The hedgehog ansatz for $SU(N)$ spherically symmetric
Skyrmions living in flat Minkowski metric
\begin{equation}
ds^{2}=-dt^{2}+dx^{2}+x^{2}(d\theta^{2}+(\sin\theta)^{2}d\varphi^{2})\ ,
\label{flat}%
\end{equation}
is based on a suitable family of projectors from $S^{2}$ into $CP^{N-1}$\ (see
\cite{su(n)3}). Such projectors $P(V)$ can be written as%

\begin{equation}
P(V)=\frac{|V\rangle\langle V|}{|V|^{2}}, \label{for}%
\end{equation}
where $|V\rangle$ is an $N$-component complex vector of two complex variables
$\xi$ and $\bar{\xi}$ which locally parametrize $S^{2}$, where $\xi
=\tan(\theta/2)\,e\sp{i\varphi}$, and $|V|^{2}=\langle V|V\rangle$ . The first
$P$ is constructed with a holomorphic $|V\rangle=|f(\xi)\rangle$ while the
Gram-Schmidt procedure gives rise to the others. Indeed, $P_{+}$ can be found
by its action on any vector $|v\rangle\in\mathbb{C}^{N}$ \cite{su(n)3} as
\begin{equation}
P_{+}|v\rangle\ = \partial_{\xi} |v\rangle\ -|v\rangle\ \,\frac{\langle
v|\,\partial_{\xi}|v\rangle}{|v|^{2}} .
\end{equation}
Consequently, the other vectors $P_{+}^{k}|v\rangle\ $ are determined
inductively: $P_{+}^{k}|v\rangle\ =P_{+}(P_{+}^{k-1}|v\rangle\ )$.

Thus the operator $P_{k}$ corresponding to the family of vectors
$|V\rangle\equiv|V_{k}\rangle=|P\sp{k}_{+}f\rangle$ (for $f=f(\xi)$) reads
\begin{equation}
P_{k}=P(P_{+}^{k}f)=\frac{|V_{k}\rangle\langle V_{k}|}{|V_{k}|^{2}}%
,\hspace{5mm}\hspace{5mm}k=0,\dots,N-1, \label{maps}%
\end{equation}
where, due to the orthogonality of the projectors, we have $\sum_{k=0}%
^{N-1}P_{k}=1$.

Due to the holomorphy of $f$, the following identities of the above defined
vectors can be proved \cite{su(n)2}:
\begin{align}
&  \langle V_{k}|V_{l}\rangle=0,\hspace{5mm}\hspace{5mm}\hspace{5mm}k\neq
l,\label{bbb}\\[3mm]
&  \partial_{\bar{\xi}}|V_{k}\rangle= -|V_{k-1}\rangle\frac{|V_{k}|^{2}%
}{|V_{k-1}|^{2}},\hspace{5mm}\hspace{5mm}\partial_{\xi}\left(  \frac
{|V_{k-1}\rangle}{|V_{k-1}|^{2}}\right)  =\frac{|V_{k}\rangle}{|V_{k-1}|^{2}}.
\label{aaa}%
\end{align}
For $SU(N)$ the components of $|V_{N-1}\rangle=P_{+}^{N-1}|f\rangle$, up to an
irrelevant overall factor which cancels in the projector, are functions of
only $\bar{\xi}$.

The $SU(N)$ hedgehog ansatz defined in \cite{su(n)1,su(n)2} reads then%
\begin{align}
U  &  =\exp\left\{  i\alpha_{0}\left(  P_{0}-\frac{\mathit{I}}{N}\right)
+i\alpha_{1}\left(  P_{1}-\frac{\mathit{I}}{N}\right)  +\dots+i\alpha
_{N-2}\left(  P_{N-2}-\frac{\mathit{I}}{N}\right)  \right\} \nonumber\\
&  =e^{-i\alpha_{0}/N}(\mathit{I}+A_{0}P_{0})\,\,e^{-i\alpha_{1}/N}%
(\mathit{I}+A_{1}P_{1})\,\dots\,\,e^{-i\alpha_{N-2}/N}(\mathit{I}%
+A_{N-2}P_{N-2})\ , \label{ansatzsu(n)}%
\end{align}
where we defined $A_{k}=e^{i\alpha_{k}}-1$. Such an ansatz involves the introduction of $N-1$ projectors and of $N-1$
profile functions $\alpha_{k}=\alpha_{k}(x)$, $k=0,\dots,N-2$; . Note that the projector $P_{N-1}$ is not included
in the above formula since it is a linear combination of the others.

One of the main results in \cite{su(n)1} and \cite{su(n)2} has been to show
that the above ansatz in Eq. (\ref{ansatzsu(n)}), when inserted into the full
Skyrme field equations in the case in which the background metric is the
standard flat metric in spherical coordinates in Eq. (\ref{flat}), gives rise
to a consistent set of $N-1$ coupled non-linear differential equations for the
$N-1$ profiles $\alpha_{i}$. Moreover, such field equations can also be
derived as stationary equations for the energy functional with respect to
variations of the profiles. A close inspection of the computations in
\cite{su(n)2} shows that the main requirement in order for the ansatz to work
is the $SO(3)$ invariance of the background metric. This then suggests that
the above ansatz may also work in the finite-volume metric in Eq.
(\ref{metric}) adopted in \cite{ccz}. This is what we show in the next section.

\section{Hedgehog ansatz for SU(N) at finite volume}

In this section we switch back to the finite-volume $SO(3)$-invariant metric
in Eq. (\ref{metric}). A direct computation shows that the ansatz in
Eq. (\ref{ansatzsu(n)}), when inserted into the full Skyrme field equations in
this metric gives rise to a consistent system of $N-1$ coupled
\textit{autonomous} non-linear differential equations for the $N-1$ profiles
$\alpha_{i}$ which is simpler than the flat \textit{non-autonomous }system
analyzed in \cite{su(n)2} as it will be now discussed. Furthermore, also in
the present case the field equations can be derived as stationary equations
for the energy functional with respect to variations of the $\alpha_{j}$'s.
It is worth emphasizing here that the explicit presence of the parameters $N$ and $R_{0}^{-2}L$ allows to consider a flat large $N$ limit in which the curvature is negligible and the product $N R_{0}^{-2}L$ remains constant. Hence, the present formalism is also relevant for people only interested in the Skyrme model on flat space-times.
It is important to notice that in such a flat limit, while the effects of the curvature of the metric in
Eq.(\ref{metric}) disappear, \textit{the effects of the cylindrical
topology do not}.

From now on we shall use the following energy $\left[  E\right]  $ and length
$\left[  L\right]  $ units:%
\begin{align}
\left[  E\right]   &  =\frac{F_{\pi}}{4e}\approx1\ GeV\ ,\label{conv1}\\
\left[  L\right]   &  =\frac{2}{eF_{\pi}}\approx0.6\ fm\ \Rightarrow
\label{conv2}%
\end{align}%
\begin{equation}
K=2\ ,\ \ \lambda=1\ . \label{conv3}%
\end{equation}

Following the same steps of \cite{su(n)2} in the new metric, one arrives at
the following expressions for the winding number, the total energy of the
$SU(N)$ hedgehog and the field equations respectively:%

\begin{equation}
W=\frac{1}{2\pi}\sum_{i=0}^{N-2}\left(  1+i\right)  \left(  N-i-1\right)
\left.  \left[  F_{i}-\sin F_{i}\right]  \right\vert _{x=-L/2}^{x=L/2}\ ,
\label{winding(n)}%
\end{equation}%
\begin{align}
E_{tot}  &  =\frac{R_{0}^{2}}{6\pi}\int\,dx\,\left[  -\frac{1}{N}\left(
\sum_{i=0}^{N-2}\overset{\cdot}{\alpha}_{i}\right)  ^{2}+\sum_{i=0}%
^{N-2}\left(  \overset{\cdot}{\alpha}_{i}\right)  ^{2}+\frac{1}{2R_{0}^{2}%
}\sum_{k=1}^{N-1}\left(  \overset{\cdot}{\alpha}_{k}-\overset{\cdot}{\alpha
}_{k-1}\right)  ^{2}D_{k}\right. \nonumber\\
&  \left.  +\frac{2}{R_{0}^{2}}\sum_{i=1}^{N-1}D_{i}+\frac{1}{4R_{0}^{4}%
}\left(  D_{1}^{2}+D_{N-1}^{2}+\sum_{k=1}^{N-2}\left(  D_{k}-D_{k+1}\right)
^{2}\right)  \right]  \ , \label{enersu(n)}%
\end{align}%
\begin{align}
0  &  =-\frac{2(l+1)}{N}\sum_{i=0}^{N-2}\left(  1+i\right)  \overset
{\cdot\cdot}{F}_{i}+\sum_{k=0}^{l}\sum_{i=k}^{N-2}\overset{\cdot\cdot}{F}%
_{i}+\frac{\left(  l+1\right)  \left(  N-l-1\right)  }{R_{0}^{2}}\left(
1-\cos\left(  F_{l}\right)  \right)  \overset{\cdot\cdot}{F}_{l}\nonumber\\
&  +\frac{\left(  l+1\right)  \left(  N-l-1\right)  }{2R_{0}^{2}}\sin\left(
F_{l}\right)  \left(  \overset{\cdot}{F}_{l}\right)  ^{2}-2\frac{\left(
l+1\right)  \left(  N-l-1\right)  }{R_{0}^{2}}\sin\left(  F_{l}\right)
\nonumber\\
&  -\frac{\left(  l+1\right)  ^{2}\left(  N-l-1\right)  ^{2}}{R_{0}^{4}%
}\left(  1-\cos\left(  F_{l}\right)  \right)  \sin\left(  F_{l}\right)
\nonumber\\
&  +\frac{\left(  l+1\right)  \left(  N-l-1\right)  \sin\left(  F_{l}\right)
}{2R_{0}^{4}}\left[  l\left(  N-l\right)  \left(  1-\cos\left(  F_{l-1}%
\right)  \right)  +\left(  l+2\right)  \left(  N-l-2\right)  \left(
1-\cos\left(  F_{l+1}\right)  \right)  \right]  \ \ \mathbf{,} \label{equ(n)}%
\end{align}
where we introduced the quantities
\begin{equation}
F_{k}=\alpha_{k}-\alpha_{k+1}\ ,\quad k=0,\ldots, N-3, \quad F_{N-2}%
=\alpha_{N-2}, \label{fkappa}%
\end{equation}
$\overset{\cdot}{F}_{l}=dF_{l}/dx$, and we defined
\begin{equation}
D_{k}=k\left(  N-k\right)  \left(  1-\cos\left(  \alpha_{k}-\alpha
_{k-1}\right)  \right) . \label{dikappa}%
\end{equation}

Notice that the winding number gets \emph{two} different contributions (while
in the Minkowski case it just gets one), as a consequence of the different
topology of the cylinder-like metric we are using. Another important point is
that when the parameter $R_{0}$ is very large the field equations
(\ref{equ(n)}) tend to linear equations, on the other hand when $R_{0}$
is small the nonlinear terms become very important. Thus, this parameter (more
precisely, the quantity $1/R_{0}^{2}$) controls the nonlinearity of the
theory, and thus it plays the r\^{o}le of an additional coupling constant. All
the analyses we shall perform in the following will confirm the fact that the
physics of the model crucially depends on it. More precisely, both the numerical analysis of Sect. \ref{NumSol} and the existence theorem of the appendix tell us that the global properties of the solutions of the system depend on the dimensionless parameter $R_0/L$, which controls the shape of the cylinder.

By comparing the present total energy in Eq. (\ref{enersu(n)}) and the field
equations in Eq. (\ref{equ(n)}) with the corresponding expressions in
\cite{su(n)2}, the first of the advantages mentioned above in working within
the metric in Eq. (\ref{metric}) is apparent. Namely, the field equations
become an autonomous system which can, as such, be analyzed with the powerful
tools of dynamical systems theory.

In the special case in which all the profiles are equal, i.e. $F_{l}=F,
\,\,\forall\,\, l$ all the equations of the above system become proportional
and equivalent to:
\begin{equation}
\label{EqprofEq}\overset{\cdot\cdot}{F}\left(  1+\frac{(1-\text{cos}%
F)}{R_{0}^{2}}\right)  - \frac{\text{sin}F}{2R_{0}^{2}}\left[  4-
\left(  \overset{\cdot}{F}^{2} - \frac{2(1-\text{cos}F)}{R_{0}^{2}}\right)
\right]  =0
\end{equation}
Upon setting $F=2\alpha$, this is precisely Eq. \eqref{SU(2)static}, valid in
the $SU(2)$ case. This is to be expected since setting all the profiles to be
equal corresponds to looking for solutions which are embeddings of the $SU(2)$ ones in $SU(N)$.

In the next sections we shall see numerically in some cases that the above
system of equations does admit nontrivial solutions, i.e. solutions whose
profiles are not equal nor proportional, while in the appendix we shall prove
analytically the existence of such solutions. Such solutions have winding
numbers which in general are nonvanishing, therefore can be interpreted as
genuine $SU(N)$ multi-Skyrmion configurations.

\subsection{The flat-large $N$ limit}
As said, the formalism developed so far strongly relies on the usage of a spacetime whose spatial sections have a finite radius $R_0$, and whose curvature goes to zero as $R_0\rightarrow\infty$. In this section we will show that there is a natural way to accommodate this flat limit, which requires to take the large $N$ limit as well. The latter is a common tool in Quantum Field Theory, see e.g. \cite{Moshe:2003xn} for a review.

The field equations in Eq. (\ref{equ(n)}) can be rewritten as follows%
\begin{align}
0  &  =\frac{1}{N}\left(  -2(l+1)\sum_{i=0}^{N-2}\left(  1+i\right)
\overset{\cdot\cdot}{F}_{i}+2N\sum_{k=0}^{l}\sum_{i=k}^{N-2}\overset
{\cdot\cdot}{F}_{i}\right)  +\frac{\left(  N-l-1\right)  }{\left(  N-1\right)
}\frac{\left(  l+1\right)  }{R^{2}}\left(  1-\cos\left(  F_{l}\right)
\right)  \overset{\cdot\cdot}{F}_{l}\nonumber\\
&  +\frac{\left(  N-l-1\right)  }{\left(  N-1\right)  }\frac{\left(
l+1\right)  }{2R^{2}}\sin\left(  F_{l}\right)  \left(  \overset{\cdot}{F}%
_{l}\right)  ^{2}-2\frac{\left(  N-l-1\right)  }{\left(  N-1\right)  }%
\frac{\left(  l+1\right)  }{R^{2}}\sin\left(  F_{l}\right) \nonumber\\
&  -\frac{\left(  N-l-1\right)  }{\left(  N-1\right)  ^{2}}^{2}\frac{\left(
l+1\right)  ^{2}}{R^{4}}\left(  1-\cos\left(  F_{l}\right)  \right)
\sin\left(  F_{l}\right) \nonumber\\
&  +\frac{\left(  N-l-1\right)  }{\left(  N-1\right)  }\frac{\left(
l+1\right)  \sin\left(  F_{l}\right)  }{2R^{4}}\left[  l\frac{\left(
N-l\right)  }{\left(  N-1\right)  }\left(  1-\cos\left(  F_{l-1}\right)
\right)  +\left(  l+2\right)  \frac{\left(  N-l-2\right)  }{\left(
N-1\right)  }\left(  1-\cos\left(  F_{l+1}\right)  \right)  \right]
\ \ \mathbf{,} \label{largeN1}%
\end{align}
where we introduced the effective radius
\begin{equation}
R^{2}=\frac{R_{0}^{2}}{N-1}\ . \label{largeN2}%
\end{equation}

The nonlinear part of the field equations (\ref{largeN1}) behaves smoothly in the large $N$ limit
provided also $R_{0}$ is large, so that $R$ is kept constant. More precisely:
\begin{equation}
N\rightarrow\infty\ ,\ R_{0}^{2}\rightarrow\infty\, \vert\ \ \underset
{N\rightarrow\infty}{\lim}\frac{R_{0}^{2}}{N-1}=R^{2} =finite\, .
\label{largeN4}%
\end{equation}
Hence, the proper way to consider the large $N$ limit is to simultaneously
consider the \emph{flat} limit $R_{0}^{2}\rightarrow\infty$ in such a way that
the parameter $R^{2}$ in Eq. (\ref{largeN2}) remains finite. Since in the
large $N$ limit in Eq. (\ref{largeN4}) the nonlinear part of the field equations Eq. (\ref{largeN1})
does not depend on $N$ and $R_{0}$ separately but only on the effective radius
$R$ defined in Eq. (\ref{largeN2}), the quantity $1/R^{2}= (N-1)/R_{0}^{2}$
plays the r\^{o}le of a \emph{geometric} (it defines a length scale) 't Hooft
coupling. Physically, the parameter $R^{2}$ in Eq. (\ref{largeN2}) represents the ``effective area" available for each Skyrmion within each section of the tube.

In conclusion, one can see the metric defined in Eq. (\ref{metric}) just as a
technical device in order to analyze multi-Skyrmionic configurations since, at
the end, one can turn off all the effects of the curvature (keeping the
cylindrical topology). In this limit, the solutions of the field equations
represent multi-Skyrmionic configuration living in a \textit{flat} tube-shaped
region whose sections have a radius much bigger than the scale $R$.

When N is large the equations read:
\begin{align}
0  &  =\frac{1}{N}\left(  -2(l+1)\sum_{i=0}^{N-2}\left(  1+i\right)
\overset{\cdot\cdot}{F}_{i}+2N\sum_{k=0}^{l}\sum_{i=k}^{N-2}\overset
{\cdot\cdot}{F}_{i}\right)  +\frac{\left(  l+1\right)  }{R^{2}}\left(
1-\cos\left(  F_{l}\right)  \right)  \overset{\cdot\cdot}{F}_{l}\nonumber\\
&  +\frac{\left(  l+1\right)  }{2R^{2}}\sin\left(  F_{l}\right)  \left(
\overset{\cdot}{F}_{l}\right)  ^{2}-2 \frac{\left(  l+1\right)  }{ R^{2}}%
\sin\left(  F_{l}\right) \nonumber\\
&  -\frac{\left(  l+1\right)  ^{2}}{R^{4}}\left(  1-\cos\left(  F_{l}\right)
\right)  \sin\left(  F_{l}\right) \nonumber\\
&  +\frac{\left(  l+1\right)  \sin\left(  F_{l}\right)  }{2R^{4}}\left[
l\left(  1-\cos\left(  F_{l-1}\right)  \right)  +\left(  l+2\right)  \left(
1-\cos\left(  F_{l+1}\right)  \right)  \right]  \ \ \mathbf{,}
\label{largeN10}%
\end{align}
The first two terms in the above equation (involving summations) describe the free part of the theory,
while the rest (involving $R$) is responsible for interactions, therefore
defines nontrivial dynamics. Therefore we find our prescription \eqref{largeN4}
a reasonable way to implement flat (large volume) limit within our formalism.
In the next section we will demonstrate
(for $N=3$), that the dynamics \emph{does} depend on $R$.

\section{Phase space portrait and geometrical 't Hooft parameter}

\label{FPoints}
One of the strongest advantages of our formalism lies in the fact that we have an
\emph{autonomous} system of differential equations. The latter can be
qualitatively described by its phase portrait, an important characteristic
of which is the set of critical points. In particular if under some change of parameters
the set of critical points remains the same (their number and their character), then
one does not expect that dynamics of the system changes dramatically under the same change.
On the contrary, if the number and/or character of the critical points changes,
one may definitely say that the dynamics of the system is qualitatively different.

In this section we are going to study our system by using the tools of
dynamical system theory (we will follow \cite{shifman, shifman2}). In particular, we will analyze how changing the value
of $R^{2}$ the number of
the critical points of the dynamical system, as well as their linear
stability properties, change.

For simplicity, we shall perform the analysis in the $SU(3)$ case. In this
case we have two profiles, $F_{0}$ and $F_{1}$, and the equations of motion are given by:
\begin{align}
\overset{{\cdot\cdot}}{F_{0}}\left(  \frac{2}{3}+\frac{1}{2R^{2}}\left(
1-\text{cos}F_{0}\right)  \right)  +\frac{1}{3}\overset{{\cdot\cdot}}{F_{1}%
}+\frac{\text{sin}F_{0}}{4R^{2}}\left[  \overset{{\cdot}}{F_{0}}^{2}%
-4-\frac{2\left(  1-\text{cos}F_{0}\right)  }{R^{2}}+\frac{\left(
1-\text{cos}F_{1}\right)  }{R^{2}}\right]   &  =0\ ,\nonumber \\
\overset{{\cdot\cdot}}{F_{1}}\left(  \frac{2}{3}+\frac{1}{2R^{2}}\left(
1-\text{cos}F_{1}\right)  \right)  +\frac{1}{3}\overset{{\cdot\cdot}}{F_{0}%
}+\frac{\text{sin}F_{1}}{4R^{2}}\left[  \overset{{\cdot}}{F_{1}}^{2}%
-4-\frac{2\left(  1-\text{cos}F_{1}\right)  }{R^{2}}+\frac{\left(
1-\text{cos}F_{0}\right)  }{R^{2}}\right]   &  =0\ . \label{eqmN3}
\end{align}
Notice that the two equations go one into the other by exchanging $F_{0}$ and
$F_{1}$. We are going to exploit this symmetry.

In order to study critical points  of the system (\ref{eqmN3}), first of all
we rewrite it as a system of first order equations as follows:
\begin{equation}
\frac{d}{dx}\vec{Z} =\vec{G}\left(\vec{Z}\right), \label{1odeq}
\end{equation}
where
\begin{equation}
\vec{Z}\equiv \left[\begin{array}{c} \Delta \\ E \\ F_0 \\F_1 \end{array}\right],
\quad \vec{G}\left(\vec{Z}\right)
= \left[\begin{array}{c} -\hat a^{-1}\left[\begin{array}{c} Q(F_0,F_1,\Delta) \\ Q(F_1,F_0,E) \end{array}\right] \\ \Delta \\E
\end{array}\right],
\end{equation}
and we introduced the following notations for brevity
\begin{equation}
\hat a \equiv \left[  \begin{array}{cc}  \frac{2}{3} + \frac{1}{2}\frac{1 - \cos{F_0}}{R^2} & \frac{1}{3} \\ \frac{1}{3} &  \frac{2}{3} + \frac{1}{2}\frac{1 - \cos{F_1}}{R^2} \end{array} \right], \quad Q(\xi, \eta,\theta) \equiv \frac{\text{sin} \xi }{4R^{2}}\left[  \theta^2 %
-4-\frac{2\left(  1-\text{cos} \xi \right)  }{R^{2}}+\frac{\left(
1-\text{cos}\eta\right)  }{R^{2}}\right].
\end{equation}

Stationary points $\vec{Z_*}$ are defined by vanishing of the vector field
$\vec{G}$, which defines the flow of the autonomous equation \eqref{1odeq} in phase space:
\begin{equation}
\vec{G}\left(\vec{Z_*}\right) = 0.
\end{equation}

One can easily check that $\det{\hat a} > 0$, therefore $\vec{Z_*}$ can be written as follows:
\begin{equation}
\vec{Z_*}  = \left[ \begin{array}{c}  0 \\0 \\F_0^* \\ F_1^*\end{array}
\right]
\end{equation}
where $F_0^*$ and $F_1^*$ are solutions of the following algebraic equations
\begin{equation}
\left\{ \begin{array}{c} Q(F_0^*,F_1^*,0) = 0 \\   Q(F_1^*,F_0^*,0) = 0 \end{array}\right. \label{cpeqs}.
\end{equation}

Equations \eqref{cpeqs} imply one of the three possibilities (the fourth possibility does not have real solutions):

\begin{enumerate}
\item $\text{sin}F_{0}=0$; $\text{sin}F_{1}=0$;

\item $\text{sin}F_{1}=0$; $-4- \frac{2\left(  1-\text{cos}F_{0}\right)
}{R^{2}}+ \frac{\left(  1-\text{cos}F_{1}\right)  }{R^{2}}=0$;

\item $\text{sin}F_{0}=0$; $-4- \frac{2\left(  1-\text{cos}F_{1}\right)
}{R^{2}}+ \frac{\left(  1-\text{cos}F_{0}\right)  }{R^{2}}=0$.
\end{enumerate}

Linearizing \eqref{1odeq} in the vicinity of the critical point $\vec{Z_*}$ we obtain
\begin{equation}
\frac{d\vec{\epsilon}}{dx}  =  J \vec{\epsilon}, \quad \vec{\epsilon} \equiv \vec{Z} - \vec{Z_*},
\end{equation}
where $J$ is the jacobian matrix of the diffeomorphism $\vec{Z}\mapsto \vec{G}(\vec{Z})$ taken at the point $\vec{Z_*}$
\begin{equation}
\hat J_{j,k} \equiv \frac{\partial {G}_{j}}{\partial Z_{k}}\bigg|_{\vec{Z} = \vec{Z_*}},\quad j,k =1,..4.
\end{equation}
In order to analyze stability of the critical point $\vec{Z_*}$ i.e. in order to understand whether phase trajectories starting
near this point ``run away" exponentially or not, one has to check whether eigenvalues of $J$ have real part or they
are purely imaginary.

Below we present a complete analysis of the critical points. Since in the algebraic equations
\eqref{cpeqs} the unknowns $F_0$ and $F_1$ enter via sine and cosine, it is sufficient to consider
critical points modulo $2\pi$.

\subsection{Case 1}

We identify the following four independent subcases:

$\bullet$ $F_{0}^{*}=0, F_{1}^{*}=0$;

$\bullet$ $F_{0}^{*}=\pi, F_{1}^{*}=\pi$;

$\bullet$ $F_{0}^{*}=0, F_{1}^{*}=\pi$;

$\bullet$ $F_{0}^{*}=\pi, F_{1}^{*}=0$.

\subsubsection{Case $F_{0}^{*}=F_{1}^{*}=0$}

In the (equal profile) case $F_{0}^{*}=F_{1}^{*}=0$ jacobian matrix is
\begin{equation}
J =
\left[ \begin {array}{cccc} 0&0&\frac{2}{R^2}&-\frac{1}{R^2}
\\ \noalign{\medskip}0&0&-\frac{1}{R^2}&\frac{2}{R^2}\\ \noalign{\medskip}1&0
&0&0\\ \noalign{\medskip}0&1&0&0\end {array} \right]
\end{equation}
whose four eigenvalues are:
\begin{equation}
\lambda_{1,2}=\pm\frac{1}{R},\qquad
\lambda_{3,4}=\pm\frac{\sqrt{3}}{R}.%
\end{equation}
Therefore this fixed point is unstable;
this property does not depend on $R$.

\subsubsection{Case $F_{0}^{*}=F_{1}^{*}=\pi$}

Also in this case we have equal profiles $F_{0}^{*}=F_{1}^{*}=\pi$. In this
case the jacobian matrix is
\begin{equation}
J = \left[ \begin {array}{cccc} 0&0&-\frac{1}{2}\,{\frac { \left( 2\,{R}^{2}+3
 \right) \left( 2\,{R}^{2}+1 \right) }{{R}^{2} \left( {R}^{4}+4\,{R}^
{2}+3 \right) }}&\frac{1}{2}\,{\frac {2\,{R}^{2}+1}{{R}^{4}+4\,{R}^{2}+3}}
\\ \noalign{\medskip}0&0&\frac{1}{2}\,{\frac {2\,{R}^{2}+1}{{R}^{4}+4\,{R}^{2}
+3}}&-\frac{1}{2}\,{\frac {4\,{R}^{4}+8\,{R}^{2}+3}{{R}^{2} \left( {R}^{4}+4\,
{R}^{2}+3 \right) }}\\ \noalign{\medskip}1&0&0&0\\ \noalign{\medskip}0
&1&0&0\end {array} \right]
\end{equation}
and the eigenvalues
\begin{equation}
\lambda_{1,2} = \pm \frac{i\sqrt {6}}{2}\,{\frac {\sqrt { \left( {R}^{2}+3 \right) \left( 2\,{R}
^{2}+1 \right) }}{ \left( {R}^{2}+3 \right) R}}, \quad \lambda_{3,4}  = \pm \frac{i\sqrt {2}}{2}\,{\frac {\sqrt { \left( {R}^{2}+1 \right) \left( 2\,{R}
^{2}+1 \right) }}{ \left( {R}^{2}+1 \right) R}}
\end{equation}
are purely imaginary, therefore this point is stable for all $R$.

\subsubsection{Case $F_{0}^{*}=0$, $F_{1}^{*}=\pi$}

Now we consider $F_{0}^{*}=0$, $F_{1}^{*}=\pi$. In this case the jacobian matrix
\begin{equation}
J=
\left[ \begin {array}{cccc} 0&0&\frac{1}{2}\,{\frac { \left( 2\,{R}^{2}-1
 \right) \left( 2\,{R}^{2}+3 \right) }{{R}^{4} \left( {R}^{2}+2
 \right) }}&{\frac {{R}^{2}+1}{{R}^{2} \left( {R}^{2}+2 \right) }}
\\ \noalign{\medskip}0&0&-\frac{1}{2}\,{\frac {2\,{R}^{2}-1}{{R}^{2} \left( {R
}^{2}+2 \right) }}&-\,{\frac {2{R}^{2}+1}{{R}^{2} \left( {R}^{2}+2
 \right) }}\\ \noalign{\medskip}1&0&0&0\\ \noalign{\medskip}0&1&0&0
\end {array} \right]
\end{equation}
has the eigenvalues
\begin{align}
\lambda_{1,2}  &  =\pm\frac{\left(  R^{2}+2\right)  \sqrt{  \sqrt{ 48
R^{8}+120 R^{6}+24 R^{4}-48 R^{2}+9}-3  }}{2 R^{2} },\nonumber\\
\lambda_{3,4}  &  =\pm\frac{i\left(  R^{2}+2\right)  \sqrt{  \sqrt{ 48
R^{8}+120 R^{6}+24 R^{4}-48 R^{2}+9}+3  }}{2 R^{2} }%
\end{align}
Unlike the previous cases, now the properties of this critical point change at a critical value of the parameter $R$.
Indeed, while the eigenvalues $\lambda_3$ and $\lambda_4$ are always purely imaginary, $\lambda_1$ and $\lambda_2$ are purely imaginary only when $R< 1/\sqrt{2}$. In this regime this critical point is stable. For $R\geq1/\sqrt{2}$ instead the first two eigenvalues become real, therefore
the fixed point becomes unstable. This is a nice example of how the stability properties change at a critical
value of the parameter. Interestingly, the value of $R$ at which the transition happens is
the same below which new fixed points appear (i.e. those occurring in cases 2 and 3 described above).

\subsubsection{Case $F_{0}^{*}=\pi$, $F_{1}^{*}=0$}

In this case the equations are exactly the same of case 3, only with $\delta$
and $\epsilon$ exchanged. The eigenvalues are therefore exactly the same so
that the same discussion applies.

\subsection{Cases 2 and 3}

We now consider cases 2 and 3 in the above list. Since they are related by the
exchange $F_{0}\leftrightarrow F_{1}$, they will have the same eigenvalues and
hence the same stability properties. Let us then consider case 2:
\begin{equation}
\sin F_{1}=0;\qquad-4- \frac{2\left(  1-\text{cos}F_{0}\right)  }{R^{2}}+
\frac{\left(  1-\text{cos}F_{1}\right)  }{R^{2}}=0.
\end{equation}
One may immediately check that a real solution only
exists for $R\leq 1/\sqrt{2}$, and it is given by
\begin{equation}
F_{1}^{*}=\pi,\qquad F_{0}^{*}=\arccos(2R^{2}),
\end{equation}
modulo $2\pi$. For $R>1/\sqrt{2}$  this critical point disappears.
The jacobian matrix reads
\begin{equation}
J =
 \left[ \begin {array}{cccc} 0&0&{\frac { \left( 8\,{R}^{8}-14\,{R}^{4
}+3 \right) \left( 2\,{R}^{2}+3 \right) }{{R}^{2} \left( 4\,{R}^{8}-
12\,{R}^{4}+9 \right) }}&-\frac{1}{2}\,{\frac {6\,{R}^{2}+3}{2\,{R}^{4}-3}}
\\ {\medskip}0&0&-{\frac {8\,{R}^{8}-14\,{R}^{4}+3}{4\,{R}^{8}
-12\,{R}^{4}+9}}&-\frac{1}{4}\,{\frac {12\,{R}^{4}-12\,{R}^{2}-9}{{R}^{2}
 \left( 2\,{R}^{4}-3 \right) }}\\ {\medskip}1&0&0&0
\\ {\medskip}0&1&0&0\end {array} \right]
\end{equation}

and its eigenvalues are:
{\small
\begin{eqnarray}
\lambda_{1,2} &=&
\pm\frac{\sqrt {2}}{4}{\frac {\sqrt { \left( 2\,{R}^{4}-3 \right) \left( 2\,{
R}^{2}+1 \right) \left( 16\,{R}^{4}+10\,{R}^{2}-3+\sqrt {256\,{R}^{8}
+896\,{R}^{6}-284\,{R}^{4}-924\,{R}^{2}+441} \right) }}{ \left( 2\,{R}
^{4}-3 \right) R}}
\\
 \lambda_{3,4} &=& \pm\frac{\sqrt {2}}{4}{\frac {\sqrt {- \left( 2\,{R}^{4}-3 \right) \left( 2\,
{R}^{2}+1 \right) \left( -16\,{R}^{4}-10\,{R}^{2}+\sqrt {256\,{R}^{8}
+896\,{R}^{6}-284\,{R}^{4}-924\,{R}^{2}+441}+3 \right) }}{ \left( 2\,{
R}^{4}-3 \right) R}} \nonumber
\end{eqnarray} }

In the whole range $R < 1/\sqrt{2}$, the first two eigenvalues are purely imaginary,
while $\lambda_3$ and $\lambda_4$ are real, therefore this critical point
is unstable. At $R=1/\sqrt{2}$, $\lambda_3$ and $\lambda_4$ vanish.

\subsection{Summary}

The analysis we just performed discloses the crucial dependence of the
dynamics of our system on the parameter $R$. In fact we saw that there is
a critical value $R=1/\sqrt{2}$ at which the nature and the number itself of the critical
points change. For $0<R \leq 1/\sqrt{2}$ and for $R>1/\sqrt{2}$ we have two \emph{qualitatively} different phase portraits, thereby different dynamics.
These are genuine finite volume effects (since the parameter $R$ represents the available surface per Baryon within the tube-shaped region) which can be disclosed only within the present framework.

\begin{figure}[htb]
\includegraphics[scale=0.4]{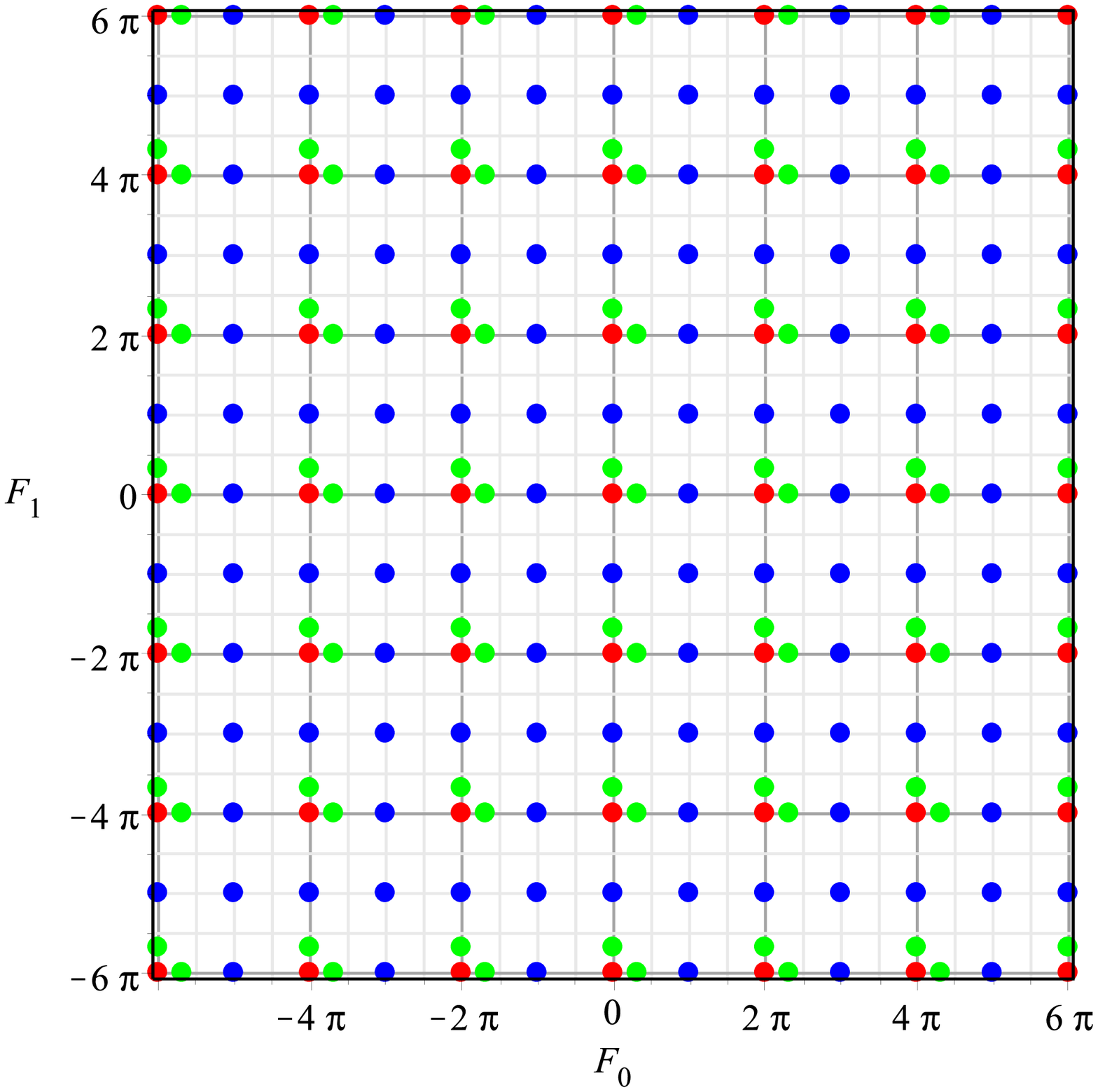}
\includegraphics[scale=0.4]{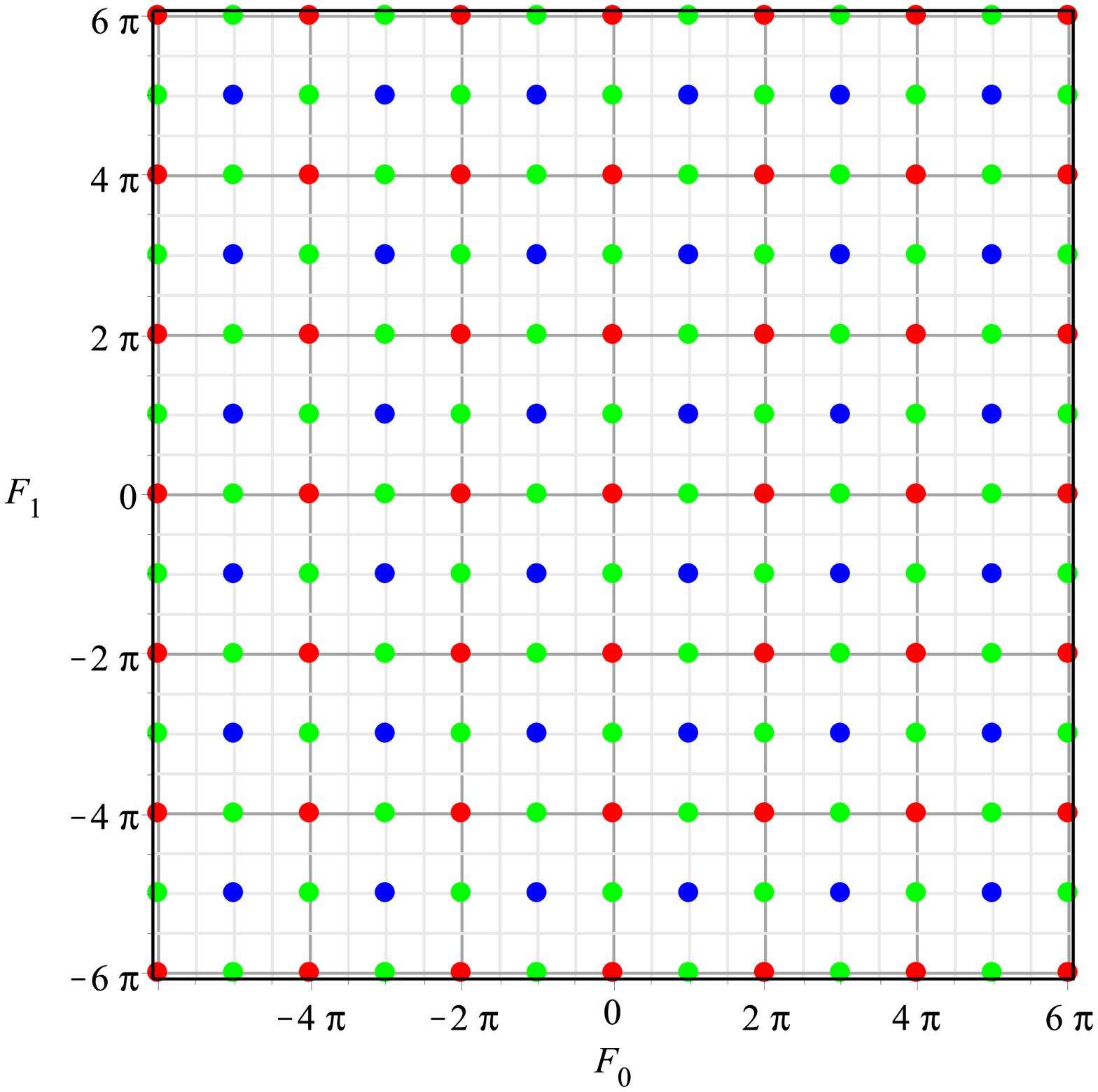}
\caption{Location of critical points for $R<1/\sqrt{2}$ on the left and for $R >1/\sqrt{2}$ on the right. In the latter case
the grid is independent on a particular choice of $R$, while for the former case we chose $R = \sqrt{\frac{\cos{1}}{2}}$. Blue points represent stable fixed points, red points represent ``absolutely" unstable fixed points (i.e. all eigenvalues of the corresponding Jacobi matrix are real),
while green points represent ``softer" unstable fixed points (i.e two eigenvalues values are real and two imaginary).  \label{F1}  }
\end{figure}

In Fig.\ref{F1} we present location of critical points for $R<1/\sqrt{2}$ and $R>1/\sqrt{2}$. As one can easily see, the two
plots are qualitatively different: while the latter grid is invariant under the discrete group of reflections $Z_2 \times Z_2$ about the axes $F_0$ and $F_1$, the former is not, that is the symmetry is broken. Moreover the former is sensitive to the particular value of $R$, while the latter is always the same for all $R>1/\sqrt{2}$. Finally, the concentration of different
critical points changes as well, since some of them changed color (i.e. type) when $R$ passed through the critical value, while new (green) fixed points appeared\footnote{For the color code see the caption.}.

These  metamorphoses of the phase portrait  hint at the presence of some sort of phase transition occurring at strong
coupling (recall that $1/R^2$ plays the r\^{o}le of a coupling constant).  It is
interesting to notice that, in our units, this critical value is of the order
of the $fm$, so although irrelevant from a flat limit perspective, it may have
important consequences in view of applications to nuclear physics, where the
dimensions of the tube-shaped region should be precisely of that order of magnitude in
order to model an atomic nucleus \cite{ccz}.

\section{Numerical Solutions: the $SU(4)$ case}
\label{NumSol}

In this section we perform a numerical study of the system in the $N=4$ case.
In this case we have three profiles, $F_{0}, F_{1}, F_{2}$, and the equations
of motion reduce to:
\begin{align*}
& \overset{{\cdot\cdot}}{F_{0}}\left( 1+\frac{2}{R_{0}^{2}}\left(
1-\text{cos}F_{0}\right) \right) + \frac{2\overset{{\cdot\cdot}}{F_{1}%
}+\overset{{\cdot\cdot}}{F_{2}}}{3} + \frac{\text{sin}F_{0}}{R_{0}^{2}}\left[
\overset{{\cdot}}{F_{0}}^{2}-4- \frac{6\left( 1-\text{cos}F_{0}\right) }%
{R_{0}^{2}}+ \frac{4\left( 1-\text{cos}F_{1}\right) }{R_{0}^{2}}\right]  =
0,\\
& \overset{{\cdot\cdot}}{F_{1}}\left( 1+\frac{2}{R_{0}^{2}}\left(
1-\text{cos}F_{1}\right) \right) + \frac{\overset{{\cdot\cdot}}{F_{0}%
}+\overset{{\cdot\cdot}}{F_{2}}}{2} +\\
& + \frac{\text{sin}F_{1}}{R_{0}^{2}}\left[ \overset{{\cdot}}{F_{1}}^{2}-4-
\frac{8\left( 1-\text{cos}F_{1}\right) }{R_{0}^{2}}+ \frac{3\left(
1-\text{cos}F_{0}\right) }{R_{0}^{2}}+ \frac{3\left( 1-\text{cos}F_{2}\right)
}{R_{0}^{2}}\right]  = 0,\\
& \overset{{\cdot\cdot}}{F_{2}}\left( 1+\frac{2}{R_{0}^{2}}\left(
1-\text{cos}F_{2}\right) \right) + \frac{\overset{{\cdot\cdot}}{F_{0}%
}+2\overset{{\cdot\cdot}}{F_{1}}}{3} + \frac{\text{sin}F_{2}}{R_{0}^{2}}\left[
\overset{{\cdot}}{F_{2}}^{2}-4- \frac{6\left( 1-\text{cos}F_{2}\right) }%
{R_{0}^{2}}+ \frac{4\left( 1-\text{cos}F_{1}\right) }{R_{0}^{2}}\right]  = 0.
\end{align*}
The first and the second go one onto the other upon exchanging $F_{0}%
\leftrightarrow F_{2}$. We can exploit this symmetry to look for solutions
which have $F_{0}=F_{2}\equiv F$ (another possibility allowed by the symmetry
is to set $F_{0}=-F_{2}$ and at the same time $F_{1}=0$, however this would
give vanishing winding number, so we do not consider it). Let us also call
$F_{1}\equiv G$. In this case we get the two independent equations:
\begin{align*}
\label{SU(4)eqs}\overset{{\cdot\cdot}}{F}\left( 1+\frac{2}{R_{0}^{2}}\left(
1-\text{cos}F\right) \right) + \frac{\overset{{\cdot\cdot}}{F}+2\overset
{{\cdot\cdot}}{G}}{3} + \frac{\text{sin}F}{R_{0}^{2}}\left[ \overset{{\cdot}%
}{F}^{2}-4- \frac{6\left( 1-\text{cos}F\right) }{R_{0}^{2}}+ \frac{4\left(
1-\text{cos}G\right) }{R_{0}^{2}}\right]  = 0,\\
\overset{{\cdot\cdot}}{G}\left( 1+\frac{2}{R_{0}^{2}}\left( 1-\text{cos}%
G\right) \right) + \overset{{\cdot\cdot}}{F}+ \frac{\text{sin}G}{R_{0}^{2}%
}\left[ \overset{{\cdot}}{G}^{2}-4- \frac{8\left( 1-\text{cos}G\right) }%
{R_{0}^{2}}+ \frac{6\left( 1-\text{cos}F\right) }{R_{0}^{2}}\right]  = 0,
\end{align*}
and the energy density (which is defined by $E=R_{0}^{2}(6\pi)^{-1}\int\mathcal{E}%
d\,x$) is given by:
\begin{align}
\mathcal{E} & = 2\overset{{\cdot}}{F}^{2} + \overset{{\cdot}}{G}^{2} +
2\overset{{\cdot}}{F}\overset{{\cdot}}{G} +\frac{1}{R_{0}^{2}}\left[ 3\left(
\overset{{\cdot}}{F}^{2}+4\right) \left( 1-\text{cos}F\right) + 2\left(
\overset{{\cdot}}{G}^{2}+4\right) \left( 1-\text{cos}G\right) \right]
+\nonumber\\
& + \frac{1}{R_{0}^{4}}\left[ 9\left( 1-\text{cos}F\right) ^{2} -12 \left(
1-\text{cos}F\right) \left( 1-\text{cos}G\right)  + 8\left( 1-\text{cos}%
G\right) ^{2}\right] .
\end{align}
Before solving the system we must choose a set of boundary conditions to
impose on $F$ and $G$ in such a way to get solutions with a nontrivial winding
number. Recall that we have to impose (anti-)periodic boundary conditions on
the field $U$:
\begin{equation}
U\left( -\frac{L}{2}\right)  = \pm\, U\left( \frac{L}{2}\right)
\end{equation}
For simplicity we shall limit ourselves to the periodic case. According to the
ansatz we are using the field can be written as:
\begin{align}
U  & =\exp\left\{ i\alpha_{0}\left( P_{0}-\frac{\mathit{I}}{4}\right)
+i\alpha_{1}\left( P_{1}-\frac{\mathit{I}}{4}\right) +i\alpha_{2}\left(
P_{2}-\frac{\mathit{I}}{4}\right) \right\} \nonumber\\
& =e^{-i\alpha_{0}/4}(\mathit{I}+A_{0}P_{0})\,\,e^{-i\alpha_{1}/4}%
(\mathit{I}+A_{1}P_{1})\,e^{-i\alpha_{2}/4}(\mathit{I}+A_{2}P_{2})\ .
\label{ansatzsu(4)}%
\end{align}
It is convenient to use the second form. We notice that the matrix $U$ is
diagonal in the basis $\{|V_{0}\rangle, |V_{1}\rangle, |V_{2}\rangle,\}$,
therefore the above boundary condition is equivalent to its diagonal matrix
elements. The diagonal matrix elements of $U$ are
\begin{align}
\langle V_{0}|U|V_{0}\rangle=e^{\frac{i}{2}\,(2F+G)},\quad\langle
V_{1}|U|V_{1}\rangle=e^{\frac{i}{2}\,G} = (\langle V_{2}|U|V_{2}\rangle)^{*}.
\end{align}
Thus the boundary conditions are satisfied if
\begin{equation}
\label{SU(4)bcs1}F\left( -\frac{L}{2}\right)  = F\left( \frac{L}{2}\right)  +
2 m \pi, \quad m\in\mathbb{Z},\qquad G\left( -\frac{L}{2}\right)  = G\left(
\frac{L}{2}\right)  + 4 n \pi, \quad n\in\mathbb{Z}.
\end{equation}
With these conditions the winding number is given by
\begin{equation}
W=6\,m + 8\,n.
\end{equation}
We also need to impose conditions on the first derivative of $U$, i.e. of $F$
and $G$. We choose the
simplest possibility which is compatible with the periodicity of the first
derivative of $U$, i.e.
\begin{equation}
\label{SU(4)bcs2}\overset{{\cdot}}{F}\left( -\frac{L}{2}\right)  =
\overset{{\cdot}}{F}\left( \frac{L}{2}\right) , \qquad\overset{{\cdot}}%
{G}\left( -\frac{L}{2}\right)  = \overset{{\cdot}}{G}\left( \frac{L}{2}\right)
.
\end{equation}

We have solved numerically the system (\ref{SU(4)eqs}) with the boundary
conditions (\ref{SU(4)bcs1}) and (\ref{SU(4)bcs2}), for some values of the
integers $m$ and $n$, taking $R_{0}=L=1$, and we computed the energies of the
solutions. The results are summarized in the tables below and in the one in
sect. \ref{Eqprofsubsect}. We have allowed $m$ and $n$ to vary in a small
range, beyond which our numerical procedure is not very accurate.

\begin{center}
\begin{tabular}
[c]%
{p{1cm}p{1cm}p{1cm}p{1cm}|p{1cm}p{1cm}p{1cm}p{1cm}|p{1cm}p{1cm}p{1cm}p{1cm}}%
$m$ & $n$ & $W$ & $E$ & $m$ & $n$ & $W$ & $E$ & $m$ & $n$ & $W$ & $E$\\\hline
-1 & 1 & 2 & 26.68 & 1 & 0 & 6 & 11.43 & 1 & 1 & 14 & 43.87\\
-2 & 2 & 4 & 100.65 & 2 & 0 & 12 & 42.41 & 2 & 2 & 28 & 168.31\\
-3 & 3 & 6 & 228.42 & 3 & 0 & 18 & 92.08 & 3 & 3 & 42 & 376.07\\
&  &  &  & 4 & 0 & 24 & 161.61 &  &  &  & \\\hline
\end{tabular}

\end{center}

The energies are exactly the same if we change sign simultaneously to $m$ and
$n$. There are sectors in which, to a very good
approximation\footnote{the accuracy is a bit lower in the last case of the
third column, most likely because of numerical errors.}, the energy grows like
the square of the modulus of the winding number, with different coefficients
in different sectors. However, this is not true in all sectors, for example:

\begin{center}
\begin{tabular}
[c]{p{1cm}p{1cm}p{1cm}p{1cm}}%
$m$ & $n$ & $W$ & $E$\\\hline
0 & 1 & 8 & 24.48\\
0 & 2 & 16 & 93.20\\\hline
\end{tabular}

\end{center}

We notice that the solutions with the smallest winding numbers are not
energetically favored, since to achieve small winding numbers both profiles
must grow going along $x$ in opposite directions, while the smallest energy
solutions are those where only one of the profile grows while the other
oscillates and stays small, i.e. \emph{either} $m$ \emph{or} $n$ are different
from zero, but not both. Notice also that increasing $n$ costs much more than
increasing $m$, despite their contributions to the winding number being not so different.

In Figs.1-6 the plots of some solutions are reported.
\begin{figure}[ptb]
\label{m-1n1}  \centering
\subfigure
{\includegraphics*[width=5.5cm]{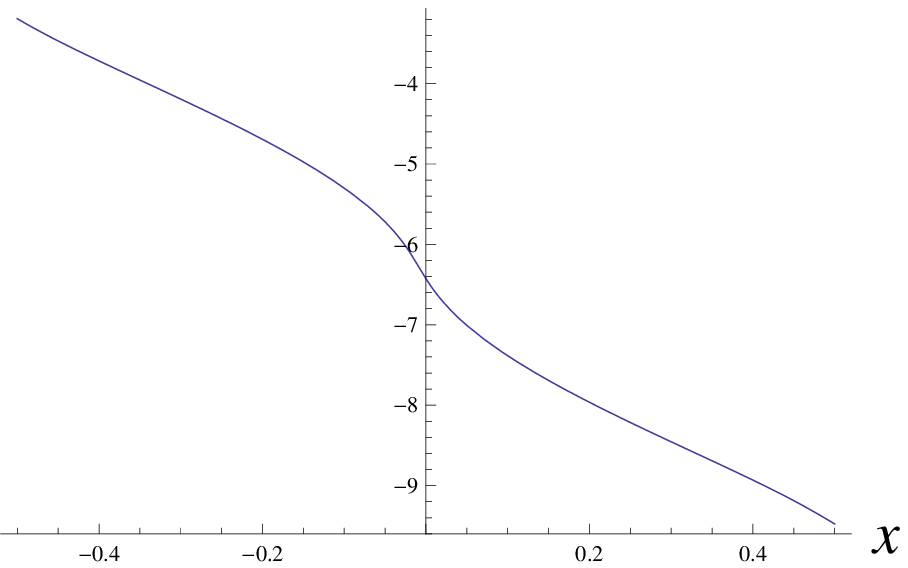}}
\subfigure
{\includegraphics*[width=5.5cm]{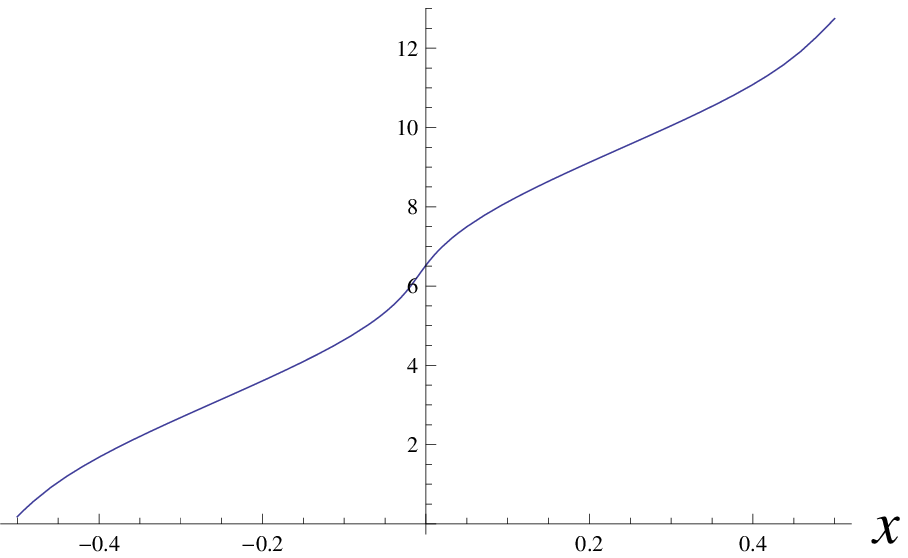}}
\subfigure
{\includegraphics*[width=5.5cm]{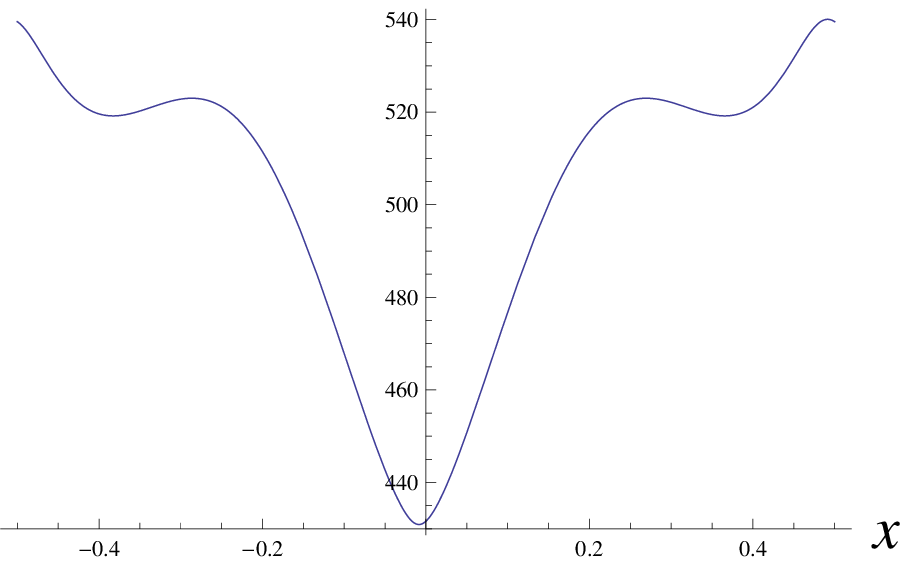}}  \caption{$F$,$G$ and
$\mathcal{E}$ for $m=-1$, $n=1$.}%
\end{figure}\begin{figure}[ptb]
\label{m-2n2}  \centering
\subfigure
{\includegraphics*[width=5.5cm]{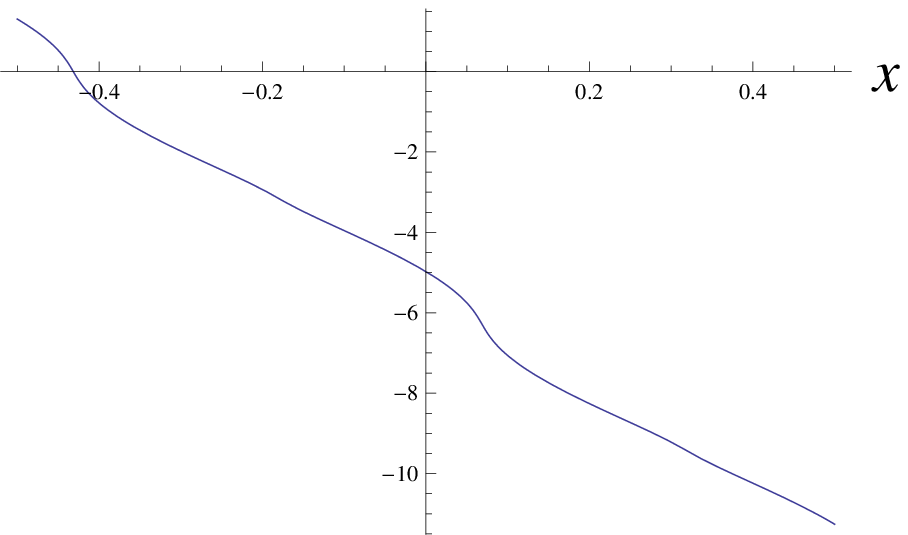}}
\subfigure
{\includegraphics*[width=5.5cm]{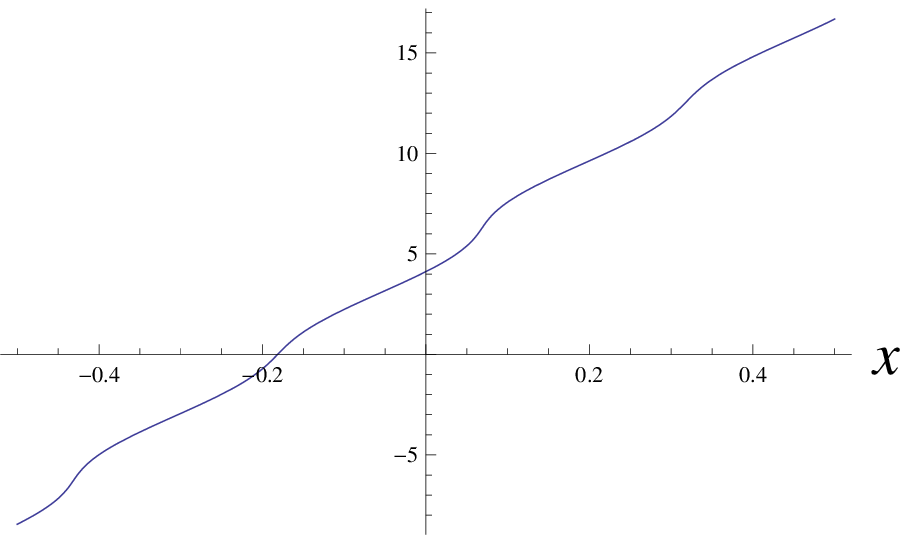}}
\subfigure
{\includegraphics*[width=5.5cm]{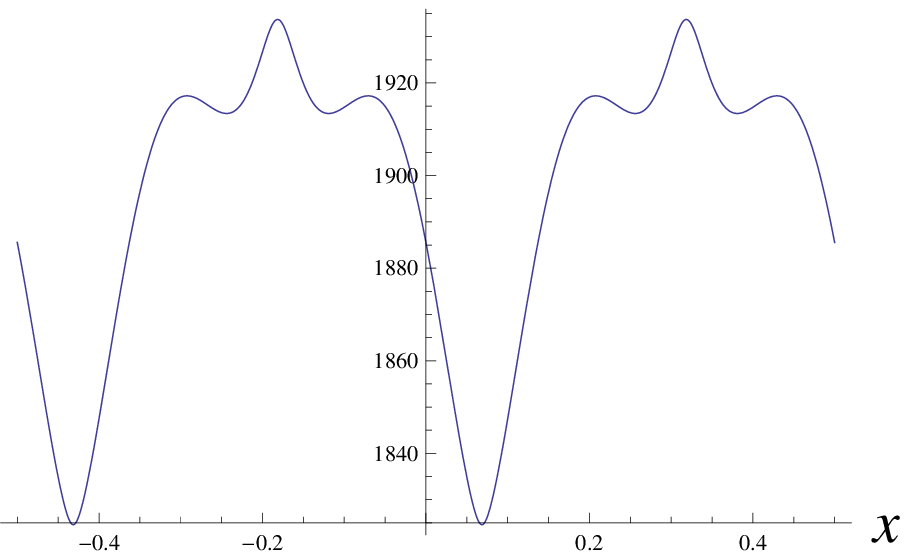}}  \caption{$F$,$G$ and
$\mathcal{E}$ for $m=-2$, $n=2$.}%
\end{figure}\begin{figure}[ptb]
\label{m1n0}  \centering
\subfigure
{\includegraphics*[width=5.5cm]{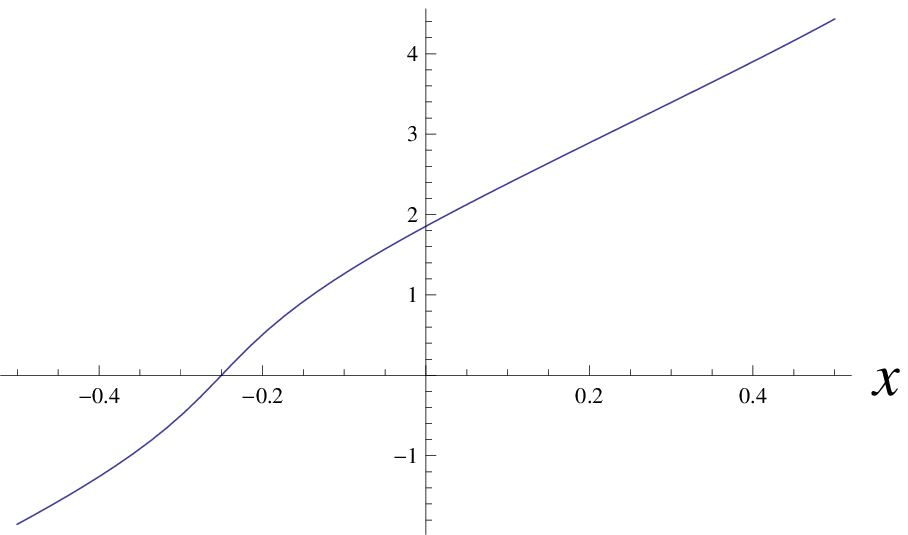}}
\subfigure
{\includegraphics*[width=5.5cm]{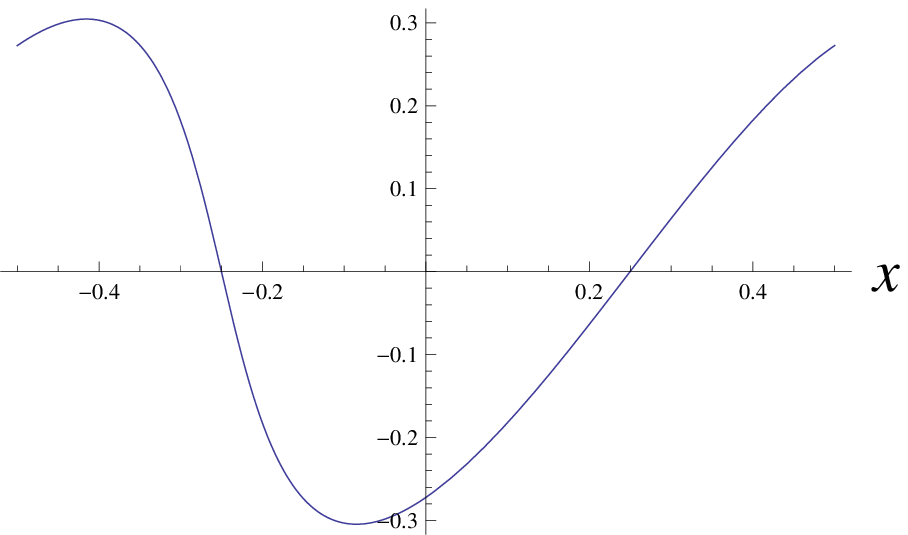}}
\subfigure
{\includegraphics*[width=5.5cm]{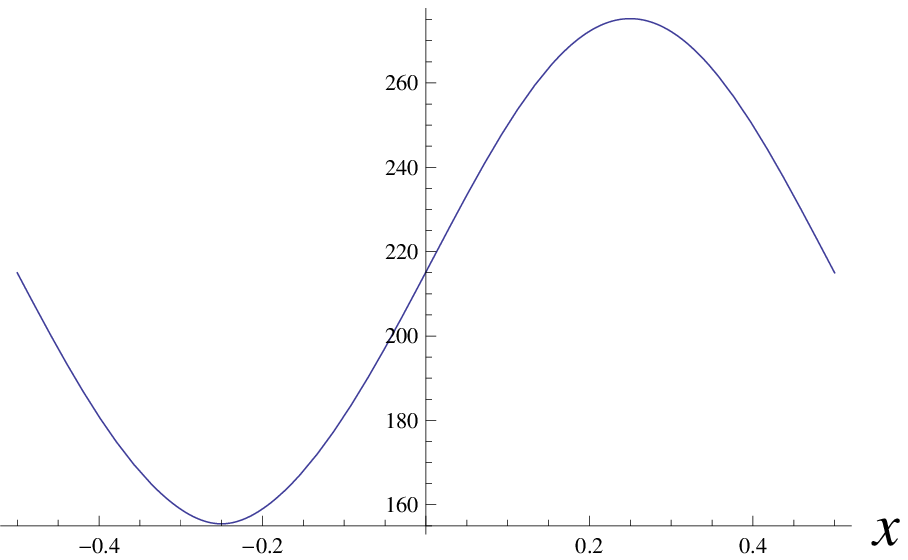}}  \caption{$F$,$G$ and
$\mathcal{E}$ for $m=1$, $n=0$.}%
\end{figure}\begin{figure}[ptb]
\label{m2n0}  \centering
\subfigure
{\includegraphics[width=5.5cm]{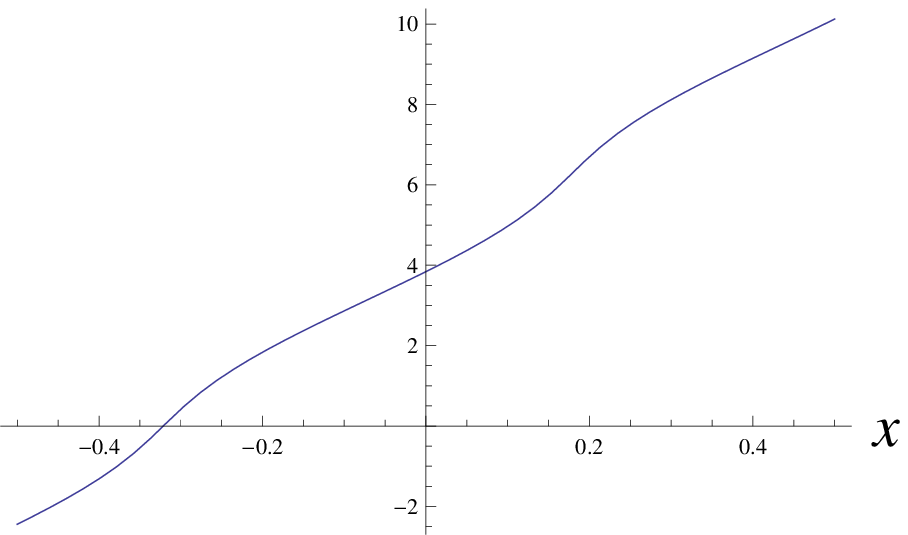}}
\subfigure
{\includegraphics[width=5.5cm]{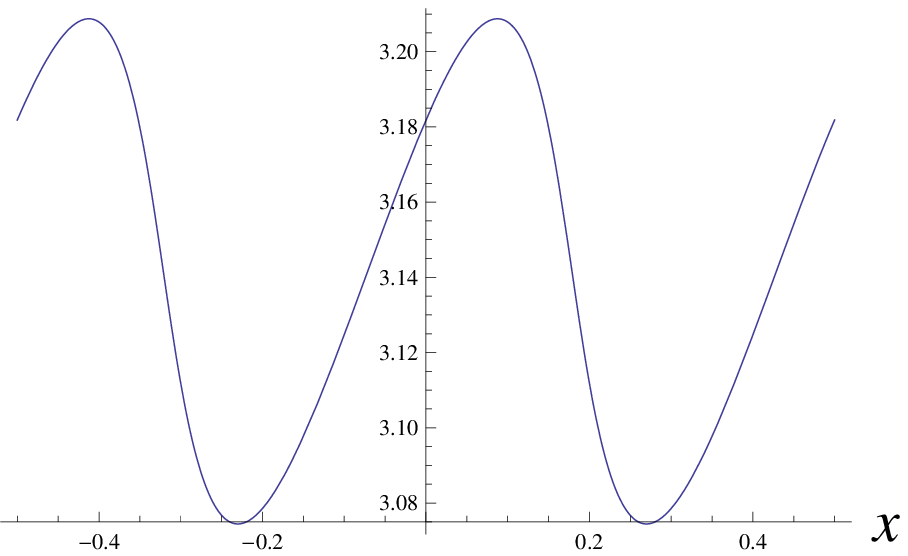}}
\subfigure
{\includegraphics[width=5.5cm]{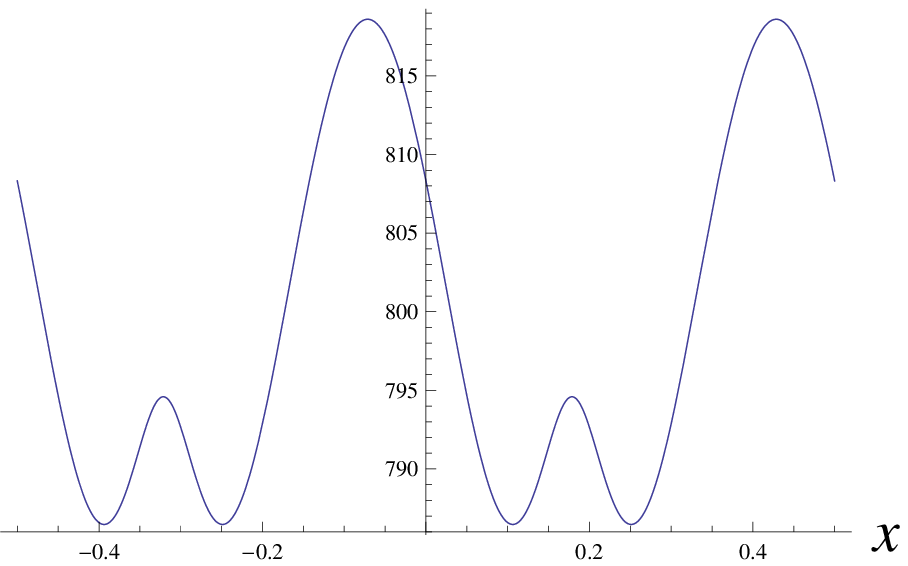}}  \caption{$F$,$G$ and
$\mathcal{E}$ for $m=2$, $n=0$.}%
\end{figure}\begin{figure}[ptb]
\label{m1n1}  \centering
\subfigure
{\includegraphics*[width=5.5cm]{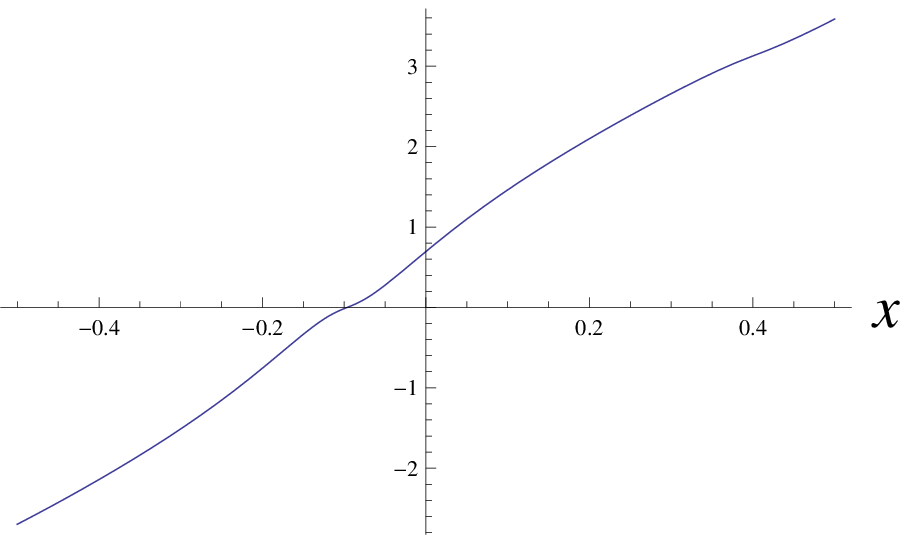}}
\subfigure
{\includegraphics*[width=5.5cm]{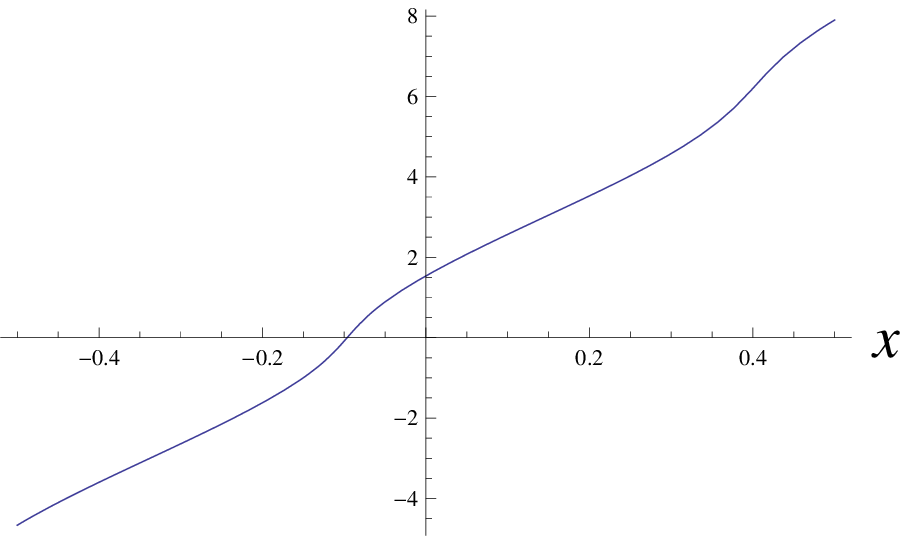}}
\subfigure
{\includegraphics*[width=5.5cm]{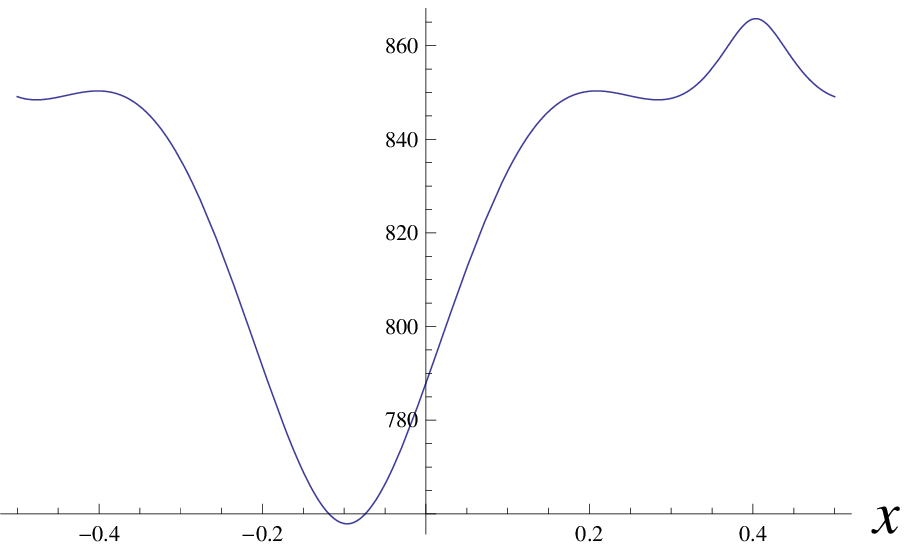}}  \caption{$F$,$G$ and
$\mathcal{E}$ for $m=1$, $n=1$.}%
\end{figure}\begin{figure}[ptb]
\label{m2n2}  \centering
\subfigure
{\includegraphics*[width=5.5cm]{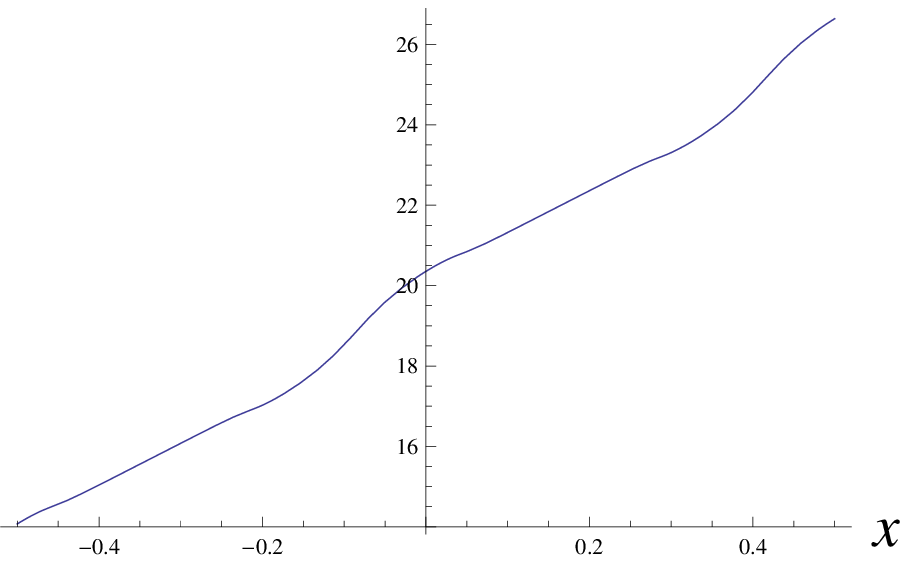}}
\subfigure
{\includegraphics*[width=5.5cm]{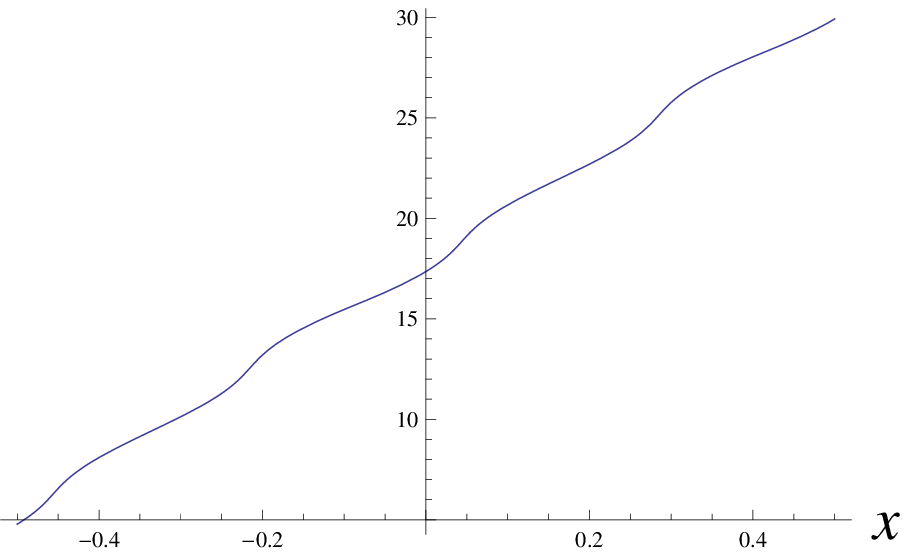}}
\subfigure
{\includegraphics*[width=5.5cm]{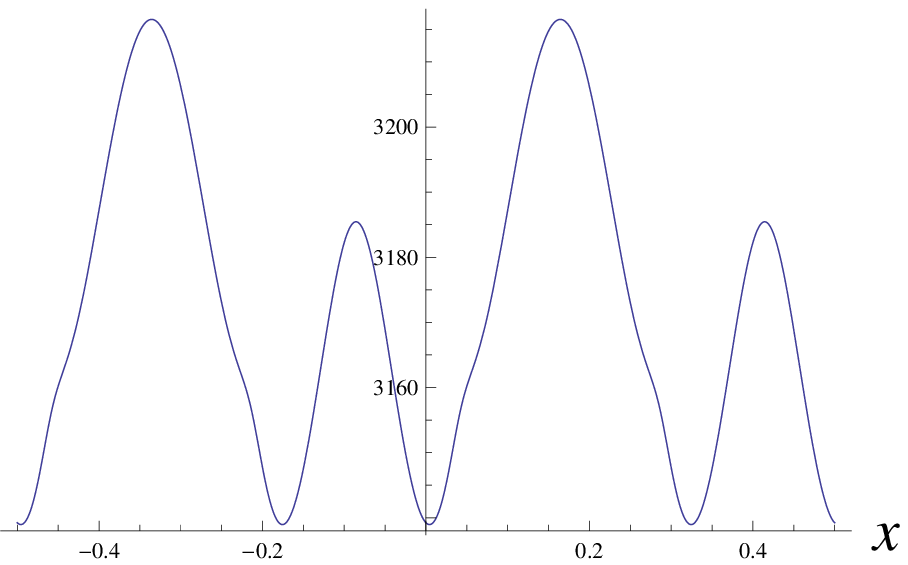}}  \caption{$F$,$G$ and
$\mathcal{E}$ for $m=2$, $n=2$.}%
\end{figure}

\subsection{Comparison with the equal profile situation}

\label{Eqprofsubsect}

If we impose that the profiles are equal, i.e $F=G\equiv F$, the equations of
motion reduce of course to (\ref{EqprofEq}). In this case the same procedure
followed above tells us that the boundary conditions to achieve a periodic $U$
are
\begin{equation}
F\left( -\frac{L}{2}\right)  = F\left( \frac{L}{2}\right)  + 4 m^{\prime}\pi,
\quad m^{\prime}\in\mathbb{Z}%
\end{equation}
in which case the winding number is simply
\begin{equation}
W_{equal}=20 \, m^{\prime}.
\end{equation}
In order to compare the equal profile situation with the non equal profile one
we have to consider cases with the same winding number. This is achieved if in
the latter case we have $m=2n$. We did the computation in the two simplest
cases, with $R_{0}=L=1$, and the results are:

\begin{center}
\begin{tabular}
[c]{p{1cm}p{1cm}p{1cm}|p{1cm}p{1cm}p{1cm}p{1cm}}%
$m^{\prime}$ & $W_{equal}$ & $E_{equal}$ & $m$ & $n$ & $W$ & $E$\\\hline
1 & 20 & 82.67 & 2 & 1 & 20 & 82.67\\
2 & 40 & 325.70 & 4 & 2 & 40 & 325.57\\\hline
\end{tabular}

\end{center}

We see that the energies are the same (in the second case the discrepancy is
due to small numerical inaccuracies: the numerical computation for the $m=4$,
$n=2$ case gave the winding number of the solutions to be 40.13, thus
signalling such inaccuracies). Therefore, in topological sectors whose winding number is compatible with equal profile configurations, with the chosen values of the parameters, equal-profile solutions and non-equal profile ones are energetically equivalent.

\subsection{Dependence on $R_{0}$}

Besides the above analysis, which was performed at fixed $R_{0}$, we also
studied what happens when this parameter is varied, considering for
definiteness the lowest winding number case, i.e $m=-1, n=1$. We checked that
when $R_{0}$ is increased the solutions tend to linear functions, in agreement
with triviality of the dynamics for large radius. Instead, for values of
$R_{0}$ (or $R_0/L$ if we allow also $L$ to vary) substantially smaller than $1$ the numerical procedure becomes
unstable and the results unreliable. This is of course to be expected since
lowering $R_{0}$ the non-linearity of the problem increases. This is in
agreement with the bounds (\ref{ban10.5}-\ref{ban11.7}), which are more
difficult to satisfy when $R_{0}$ is small, and trivial when $R_{0}$ is large.

In the small $R_{0}$ case a more sophisticated numerical analysis would therefore be
needed. In particular it would be interesting to see whether at strong
coupling non-equal profile solutions are energetically favored with respect to
equal profile ones with the same winding number, as suggested by the fixed point analysis of section
\ref{FPoints}. Another interesting issue requiring a more accurate numerical work is the identification of a range of the parameters in which the dependence of the energy on the winding number is linear rather than quadratic. We hope to come back on these interesting issues in a future investigation.

\section{Conclusions}

In this paper the $SU(N)$ Skyrme model is studied in a tube-shaped
geometry which allows to consider finite volume situations. The use of such a
geometry allows to introduce a parameter which regulates the dynamics as a new
(inverse) coupling constant. These simplifications, which
make the system autonomous, allow us to perform a detailed study of the system
using several techniques, i.e. dynamical systems, numerical solutions and
rigorous existence theorems. All these analyses confirm the relevance of the
parameter $R_{0}$ introduced by the metric on the dynamics of the system. As a
further bonus, when combined with $N$ as $1/R^{2}=N/(R_{0})^{2}$, this parameter defines an effective 't Hooft
coupling. In the large $N$ limit in which such coupling is kept fixed, flat space-time
is recovered. This effective 't Hooft parameter determines the discrete symmetries of the Skyrmions configurations.
These results clearly show that the present framework is a very promising tool to study interacting Skyrmions systems. A very interesting problem which we hope to analyse in a future publication is to extend the present analysis at finite temperature and/or chemical potential.

\subsection*{Acknowledgements}

The authors would like to thank J. Zanelli for enlightening discussions and
comments. This work has been funded by the Fondecyt grant 1120352. The Centro de Estudios Cient\'{\i}ficos (CECs) is funded by
the Chilean Government through the Centers of Excellence Base Financing
Program of Conicyt. M. K. acknowledges partial support from UniNA and Compagnia di San Paolo in the framework
of the program STAR 2013.

\section*{Appendix: Existence of Solutions}

In this appendix we prove analytically that non-trivial solutions which are
not trivial embeddings of $SU(2)$ into $SU(N)$ (in which the profiles are not
proportional) do indeed exist. We will focus for simplicity on the $SU(3)$
case but the same argument can be easily extended to the general case. The
basic mathematical tool of this section is a well-known result of nonlinear
functional analysis, the \textit{Schauder theorem} (see for a detailed
pedagogical review on mathematical tools to deal with non-linear partial
differential equations \cite{nonlinearanal}).

The statement of the Schauder theorem (\cite{nonlinearanal,nonlinearanal2}) is
the following. Let $S$ be a Banach space\footnote{A Banach space is a linear
space endowed with a norm, and which is complete with respect to the metric
induced by the norm. Recall that a metric space is a space in which a distance
$d\left(  X,Y\right)  $ between any pair of elements of the space is defined,
and it is called complete if (with respect to the chosen metric) from every
Cauchy sequence one can extract a convergent subsequence (see, for instance,
\cite{nonlinearanal}).}. Let $C$ a bounded closed convex set in $S$, and let
$T$ be a compact operator\footnote{An operator $T$ from a Banach space $S$
into itself (see, for a detailed discussion,
\cite{nonlinearanal,nonlinearanal2}) is called \textit{compact} if and only
if, for any bounded sequence $\left\{  x_{n}\right\}  $, the sequence
$\left\{  T(x_{n})\right\}  $ has a convergent subsequence.} from the Banach
space $S$ into itself such that $T$ maps $C$ into itself:%
\begin{equation}
T\left[  .\right]  :C\rightarrow C\ . \label{ban2}%
\end{equation}
Then the map $T\left[  .\right]  $ has (at least) one fixed point in $C$. In
other words, under the above hypotheses, there exists a solution to the
equation
\begin{equation}
T\left[  X\right]  =X\ . \label{ban3}%
\end{equation}

Let us recall the equations of motion for the $SU(3)$ case:
\begin{align}
\overset{{\cdot\cdot}}{F_{0}}\left(  \frac{2}{3}+\frac{1}{2R^{2}}\left(
1-\text{cos}F_{0}\right)  \right)  +\frac{1}{3}\overset{{\cdot\cdot}}{F_{1}%
}+\frac{\text{sin}F_{0}}{4R^{2}}\left[  \overset{{\cdot}}{F_{0}}^{2}%
-4-\frac{2\left(  1-\text{cos}F_{0}\right)  }{R^{2}}+\frac{\left(
1-\text{cos}F_{1}\right)  }{R^{2}}\right]   &  =0\ ,\label{SU(3)eq1}\\
\overset{{\cdot\cdot}}{F_{1}}\left(  \frac{2}{3}+\frac{1}{2R^{2}}\left(
1-\text{cos}F_{1}\right)  \right)  +\frac{1}{3}\overset{{\cdot\cdot}}{F_{0}%
}+\frac{\text{sin}F_{1}}{4R^{2}}\left[  \overset{{\cdot}}{F_{1}}^{2}%
-4-\frac{2\left(  1-\text{cos}F_{1}\right)  }{R^{2}}+\frac{\left(
1-\text{cos}F_{0}\right)  }{R^{2}}\right]   &  =0\ . \label{SU(3)eq2}%
\end{align}
We want to prove that the system in Eqs. (\ref{SU(3)eq1}) and (\ref{SU(3)eq2})
admits non-trivial solutions in which the profiles $F_{0}$ and $F_{1}$ are not
proportional. In order to achieve this goal, let us rewrite it as coupled
integral equations (for notational simplicity in this section we shall
consider $x\in[0,L]$ instead of $x\in[-L/2,L/2]$, since everything just
depends on the length $L$ of the tube):

\begin{equation}
F_{0}\left(  x\right)  =a_{0}+b_{0}x-\int_{0}^{x}H_{0}\left(  s\right)  ds\ ,
\label{int1}%
\end{equation}%
\begin{equation}
F_{1}\left(  x\right)  =a_{1}+b_{1}x-\int_{0}^{x}H_{1}\left(  s\right)  ds\ ,
\label{int2}%
\end{equation}
where
\begin{align}
H_{0}\left(  s\right)   &  =\,\int_{0}^{s}\, A_{0}\left\{  \frac
{\text{sin}F_{0}\left(  \rho\right)  }{4R^{2}}h_{0}\left(  \rho\right)
+\frac{1}{3}\overset{{\cdot\cdot}}{F_{1}}\left(  \rho\right)  \right\}
d\rho\ ,\label{int1.25}\\
H_{1}\left(  s\right)   &  =\,\int_{0}^{s}\,A_{1}\left\{  \frac{\text{sin}%
F_{1}\left(  \rho\right)  }{4R^{2}}h_{1}\left(  \rho\right)  +\frac{1}%
{3}\overset{{\cdot\cdot}}{F_{0}}\left(  \rho\right)  \right\}  d\rho
\ ,\label{int2.25}\\
h_{0}\left(  \rho\right)   &  =\left[  \overset{{\cdot}}{F_{0}}^{2}%
-4-\frac{2\left(  1-\text{cos}F_{0}\left(  \rho\right)  \right)  }{R^{2}%
}+\frac{\left(  1-\text{cos}F_{1}\left(  \rho\right)  \right)  }{R^{2}%
}\right]  \ ,\ \label{int1.35}\\
h_{1}\left(  \rho\right)   &  =\left[  \overset{{\cdot}}{F_{1}}^{2}%
-4-\frac{2\left(  1-\text{cos}F_{1}\left(  \rho\right)  \right)  }{R^{2}%
}+\frac{\left(  1-\text{cos}F_{0}\left(  \rho\right)  \right)  }{R^{2}}\right]
, \label{int2.35}\\
A_{i}  & =\left( \frac{2}{3} + \frac{1}{2\,R^{2}}\left( 1-\cos\,F_{i}\right)
\right) ^{-1}, \quad i=0,1
\end{align}
where $\overset{{\cdot}}{F_{i}}=\frac{dF_{}}{d\rho}$, $i=0,1$, and $a_{i}$ and
$b_{i}$ represent the initial data for the two profiles $F_{0}$ and $F_{1}$
and their derivatives at $x=0$. It is a trivial computation to show that the
above system of integral equation is equivalent to the system in Eqs.
(\ref{SU(3)eq1}) and (\ref{SU(3)eq2}). The system in Eqs. (\ref{int1}) and
(\ref{int2}) can be written as a fixed point condition for the following
vectorial operator $\overrightarrow{T}$ acting component-wise on pairs of
$C^{2}$ functions $\overrightarrow{F}\left(  x\right)  =\left(  F_{0}\left(
x\right)  ,F_{1}\left(  x\right)  \right) $:%
\begin{align*}
\overrightarrow{T}  &  :C^{2}\left[  0,L\right]  \times C^{2}\left[
0,L\right]  \rightarrow C^{2}\left[  0,L\right]  \times C^{2}\left[
0,L\right]  \ ,\\
\overrightarrow{F}\left(  x\right)   &  =\left(  F_{0}\left(  x\right)
,F_{1}\left(  x\right)  \right)  \in C^{2}\left[  0,L\right]  \times
C^{2}\left[  0,L\right]  \ ,
\end{align*}%
\begin{equation}
\overrightarrow{T}\left[  F_{0},F_{1}\right]  =\overrightarrow{T}\left[
\ \overrightarrow{F}\left(  x\right)  \right]  =\left(  T_{0}\left(  x\right)
,T_{1}\left(  x\right)  \right)  \ , \label{int2.5}%
\end{equation}
with
\begin{align}
T_{0}\left(  x\right)   &  =a_{0}+b_{0}x-\int_{0}^{x}H_{0}\left(  s\right)
ds\ \ ,\label{int3}\\
T_{1}\left(  x\right)   &  =a_{1}+b_{1}x-\int_{0}^{x}H_{1}\left(  s\right)
ds\ \ , \label{int4}%
\end{align}
where the functions $H_{i}\left(  s\right)  $ are defined in Eqs.
(\ref{int1.25}), (\ref{int2.25}), (\ref{int1.35}) and (\ref{int2.35}). It is
then obvious that the system in Eqs. (\ref{int1}) and (\ref{int2}) can be
written as the following fixed-point condition%
\begin{equation}
\overrightarrow{F}\left(  x\right)  =\overrightarrow{T}\left[
\ \overrightarrow{F}\left(  x\right)  \right]  \ , \label{fixedpoint}%
\end{equation}
where the operator $\overrightarrow{T}$ has been defined in Eqs.
(\ref{int2.5}), (\ref{int3}) and (\ref{int4}). Hence, now the task is to
define $\overrightarrow{T}$ as a compact operator from a bounded closed convex
sub-set of a Banach space into itself.

In order to achieve this goal, first of all let us define the following metric
into the space $C^{2}\left[  0,L\right]  \times C^{2}\left[  0,L\right]  $:%
\begin{align}
d\left(  \overrightarrow{F},\overrightarrow{G}\right)   &  =\underset
{x\in\left[  0,L\right]  }{\sup}\left\vert F_{0}(x)-G_{0}(x)\right\vert
+\underset{x\in\left[  0,L\right]  }{\sup}\left\vert \frac{dF_{0}}{dx}%
-\frac{dG_{0}}{dx}\right\vert +\underset{x\in\left[  0,L\right]  }{\sup
}\left\vert \frac{d^{2}F_{0}}{dx^{2}}-\frac{d^{2}G_{0}}{dx^{2}}\right\vert
+\nonumber\\
&  +\underset{x\in\left[  0,L\right]  }{\sup}\left\vert F_{1}(x)-G_{1}%
(x)\right\vert +\underset{x\in\left[  0,L\right]  }{\sup}\left\vert
\frac{dF_{1}}{dx}-\frac{dG_{1}}{dx}\right\vert +\underset{x\in\left[
0,L\right]  }{\sup}\left\vert \frac{d^{2}F_{1}}{dx^{2}}-\frac{d^{2}G_{1}%
}{dx^{2}}\right\vert \ ,\label{ban4.01}\\
\overrightarrow{F}\left(  x\right)   &  =\left(  F_{0}\left(  x\right)
,F_{1}\left(  x\right)  \right)  \ ,\ \overrightarrow{G}\left(  x\right)
=\left(  G_{0}\left(  x\right)  ,G_{1}\left(  x\right)  \right)  \ \in
C^{2}\left[  0,L\right]  \times C^{2}\left[  0,L\right]  \ .\nonumber
\end{align}
With respect to this metric, which is induced by a norm, the space
$C^{2}\left[  0,L\right]  \times C^{2}\left[  0,L\right]  $ is a Banach space
which we will call $S$.

\emph{The next task} to apply the Schauder theorem is to define a bounded
closed convex sub-set $C$ of the Banach space defined above (using the metric
in Eq. (\ref{ban4.01})) such that $\overrightarrow{T}$ maps $C$ into itself.
Let us define $C$ as%
\begin{align}
C  &  \equiv\left\{  \overrightarrow{F}\left(  x\right)  =\left(  F_{0}\left(
x\right)  ,F_{1}\left(  x\right)  \right)  \in\left.  S\right\vert \ \forall
x\in\left[  0,L\right]  \ \right.  \ \left\vert F_{0}(x)-a_{0}\right\vert \leq
B\ ,\ \left\vert F_{1}(x)-a_{1}\right\vert \leq B\ \ ,\label{ban5}\\
&  \left.  \ \left\vert \frac{dF_{0}}{dx}\right\vert \leq B\ ,\ \ \left\vert
\frac{d^{2}F_{0}}{dx^{2}}\right\vert \leq B\ ,\ \ \left\vert \frac{dF_{1}}%
{dx}\right\vert \leq B\ ,\ \left\vert \frac{d^{2}F_{1}}{dx^{2}}\right\vert
\leq B\right\}  \ , \quad\ \ \ B\in%
\mathbb{R}
_{+}\ ,\nonumber
\end{align}
where $a_{i}$ are the initial data appearing in Eqs. (\ref{int1}) and
(\ref{int2}) so that $C$ is closed by definition. It is easy to see that $C$
is bounded as well, since from the definition of $C$ it follows that%
\begin{align}
\left\vert F_{0}(x)\right\vert  &  \leq B+\left\vert a_{0}\right\vert
\ ,\ \left\vert \frac{dF_{0}}{dx}\right\vert \leq B\ ,\ \left\vert \frac
{d^{2}F_{0}}{dx^{2}}\right\vert \leq B\ ,\label{ban6}\\
\left\vert F_{1}(x)\right\vert  &  \leq B+\left\vert a_{1}\right\vert
\ ,\ \left\vert \frac{dF_{1}}{dx}\right\vert \leq B\ ,\ \left\vert \frac
{d^{2}F_{1}}{dx^{2}}\right\vert \leq B\ , \label{ban7}%
\end{align}
which clearly implies that there exists $M$ such that $d(\overrightarrow
{F},\overrightarrow{G})\leq M$, given any two functions $\overrightarrow
{F},\overrightarrow{G}\in C$. It remains to prove that $C$ is convex, i.e. to
check that if $\overrightarrow{F}\left(  x\right)  $ and $\overrightarrow
{G}\left(  x\right)  $ both belong to $C$ then also $\theta\overrightarrow
{F}\left(  x\right)  +\left(  1-\theta\right)  \overrightarrow{G}\left(
x\right)  $ belongs to $C$, $\forall\,\,\theta\in\left[  0,1\right]  $. This
can be seen as follows:%
\begin{align}
\left\vert \theta F_{1}\left(  x\right)  +\left(  1-\theta\right)
G_{1}\left(  x\right)  -a_{1}\right\vert  &  \leq\left\vert \theta\left(
F_{1}\left(  x\right)  -a_{1}\right)  \right\vert +\left\vert \left(
1-\theta\right)  \left(  G_{1}\left(  x\right)  -a_{1}\right)  \right\vert
\leq\theta B+\left(  1-\theta\right)  B\leq B\ ,\label{ban8}\\
\left\vert \theta F_{0}\left(  x\right)  +\left(  1-\theta\right)
G_{0}\left(  x\right)  -a_{0}\right\vert  &  \leq\left\vert \theta\left(
F_{0}\left(  x\right)  -a_{0}\right)  \right\vert +\left\vert \left(
1-\theta\right)  \left(  G_{0}\left(  x\right)  -a_{0}\right)  \right\vert
\leq\theta B+\left(  1-\theta\right)  B\leq B\ , \label{ban9}%
\end{align}
while the conditions on the derivatives are trivially satisfied.

Thus, $C$ is closed, bounded and convex. The requirement that $\overrightarrow
{T}$ maps $C$ into itself will impose some constraints on the parameters, as
we shall see shortly.

From the definition of $C$ in Eq. (\ref{ban5}) we deduce the inequalities
\begin{align}
\underset{x\in\left[  0,L\right]  }{\sup}\left\vert H_{0}\left(  x\right)
\right\vert  &  \leq L\left\{  \frac{B}{2} +\frac{3}{8R^{2}}\left[
B^{2}+4+\frac{6}{R^{2}}\right]  \right\}  \ ,\label{useful1}\\
\underset{x\in\left[  0,L\right]  }{\sup}\left\vert H_{1}\left(  x\right)
\right\vert  &  \leq L\left\{  \frac{B}{2} +\frac{3}{8R^{2}}\left[
B^{2}+4+\frac{6}{R^{2}}\right]  \right\}  \ , \label{useful2}%
\end{align}
\begin{align}
\underset{x\in\left[  0,L\right]  }{\sup}\left\vert \frac{dH_{0}}%
{dx}\right\vert  &  \leq\frac{B}{2} +\frac{3}{8R^{2}}\left[  B^{2}+4+\frac
{6}{R^{2}}\right]  \ ,\label{useful3}\\
\underset{x\in\left[  0,L\right]  }{\sup}\left\vert \frac{dH_{1}}%
{dx}\right\vert  &  \leq\frac{B}{2} +\frac{3}{8R^{2}}\left[  B^{2}+4+\frac
{6}{R^{2}}\right]  \ , \label{useful4}%
\end{align}
(where we used the fact that $A_{i}\leq3/2$), which will be needed later.

\emph{The next task} to apply the Schauder theorem is to prove that
$\overrightarrow{T}$ is a compact operator. To do this we advocate the
\textit{Ascoli-Arzel\`{a} theorem}, which states that if a sequence of
functions (defined on a compact metric space) $\left\{ f_{n}\right\} $ is
uniformly bounded and equicontinuous, then a convergent subsequence can be
extracted from it\footnote{Recall that the sequence $\left\{ f_{n}\right\} $
is uniformly bounded if $\left| f_{n}\right| <M$, where $M$ does not depend on
$n$; $\left\{ f_{n}\right\}  $ is said to be equicontinuous if, given
$\epsilon>0$, $\exists\ \delta>0$ such that $\left\vert f_{n}\left(  x\right)
-f_{n}\left(  y\right)  \right\vert <\epsilon$ whenever $\left\vert
x-y\right\vert <\delta$ \textit{and, moreover, }$\delta$\textit{ does not
depend on} $n$ (otherwise the sequence would be continuous but not
equicontinuous: see \cite{nonlinearanal}).}. Thus compactness of
$\overrightarrow{T}$ is equivalent to the statement that if $\overrightarrow
{F}_{n}\left(  x\right)  =\left(  F_{0}^{n}\left(  x\right)  ,F_{1}^{n}\left(
x\right)  \right)  $ is a sequence in $C$, then the sequence $\overrightarrow
{T}\left[  \ \overrightarrow{F}_{n}\left(  x\right)  \right]  =\left(
T_{0}\left(  n;x\right)  ,T_{1}\left(  n;x\right)  \right)  $ is uniformly
bounded and equicontinuous.

Uniform boundedness can be proved as follows:%
\[
\left\vert T_{0}\left(  n;x\right)  \right\vert =\left\vert a_{0}+b_{0}%
x-\int_{0}^{x}H_{0}\left(  n;s\right)  ds\right\vert \leq
\left\vert a_{0}\right\vert +\left\vert b_{0}L\right\vert +L\underset
{x\in\left[  0,L\right]  }{\sup}\left\vert H_{0}\left(  n;x\right)
\right\vert \ .
\]
and therefore, from Eq. (\ref{useful1}), one gets:%
\begin{equation}
\left\vert T_{0}\left(  n;x\right)  \right\vert \leq\left\vert a_{0}%
\right\vert +\left\vert b_{0}L\right\vert +L^{2}\left\{  \frac{B}{2} +\frac
{3}{8R^{2}}\left[  B^{2}+4+\frac{6}{R^{2}}\right]  \right\}  \ , \label{ban10}%
\end{equation}
and, similarly, for $\left\vert T_{1}\left(  n;x\right)  \right\vert $:
\begin{equation}
\left\vert T_{1}\left(  n;x\right)  \right\vert \leq\left\vert a_{1}%
\right\vert +\left\vert b_{1}L\right\vert +L^{2}\left\{  \frac{B}{2} +\frac
{3}{8R^{2}}\left[  B^{2}+4+\frac{6}{R^{2}}\right]  \right\}  \ . \label{ban11}%
\end{equation}

Now, the sequence of images $\overrightarrow{T}\left[  \ \overrightarrow
{F}_{n}\left(  x\right)  \right]  $ has to belong to $C$ as well. As it always
happens (see \cite{nonlinearanal} and \cite{nonlinearanal2}) this will give
some constraints on the range of the initial data as well as on the parameters
$B$, $L$ and $R$ since, from the definition of $C$, one has to require that
for any $n$
\begin{align}
\left\vert T_{0}\left(  n;x\right)  -a_{0}\right\vert  &  \leq B\ ,~\left\vert
\frac{dT_{0}\left(  n;x\right) }{dx}\right\vert =\left\vert b_{0}-H_{0}\left(
n;x\right)  \right\vert \leq B\ ,\ \left\vert \frac{d^{2}T_{0}\left(
n;x\right) }{dx^{2}}\right\vert =\left\vert \frac{dH_{0}\left(  n;x\right)
}{dx}\right\vert \leq B\ ,\label{cons1}\\
\left\vert T_{1}\left(  n;x\right)  -a_{1}\right\vert  &  \leq B,~\left\vert
\frac{dT_{1}\left(  n;x\right) }{dx}\right\vert =\left\vert b_{1}-H_{1}\left(
n;x\right)  \right\vert \leq B\ ,\ \left\vert \frac{d^{2}T_{1}\left(
n;x\right) }{dx^{2}}\right\vert =\left\vert \frac{dH_{1}\left(  n;x\right)
}{dx}\right\vert \leq B\ . \label{cons2}%
\end{align}
Consequently, as it can be easily seen comparing Eqs. (\ref{cons1}),
(\ref{cons2}), (\ref{ban5}), (\ref{ban6}) and (\ref{ban7}) with Eqs.
(\ref{useful1}), (\ref{useful2}), (\ref{useful3}), (\ref{useful4}),
(\ref{ban10}) and (\ref{ban11}), the following constraints arise:%
\begin{align}
\left\vert b_{0}L\right\vert +L^{2}\left\{  \frac{B}{2} +\frac{3}{8R^{2}%
}\left[  B^{2}+4+\frac{6}{R^{2}}\right]  \right\}   &  \leq
B\ ,\label{ban10.5}\\
\left\vert b_{1}L\right\vert +L^{2}\left\{  \frac{B}{2} +\frac{3}{8R^{2}%
}\left[  B^{2}+4+\frac{6}{R^{2}}\right]  \right\}   &  \leq B\ ,
\label{ban11.5}%
\end{align}%
\begin{align}
\left\vert b_{0}\right\vert +L\left\{  \frac{B}{2} +\frac{3}{8R^{2}}\left[
B^{2}+4+\frac{6}{R^{2}}\right]  \right\}   &  \leq B\ ,\label{ban10.6}\\
\left\vert b_{1}\right\vert +L\left\{  \frac{B}{2} +\frac{3}{8R^{2}}\left[
B^{2}+4+\frac{6}{R^{2}}\right]  \right\}   &  \leq B\ , \label{ban11.6}%
\end{align}%
\begin{align}
\frac{B}{2} +\frac{3}{8R^{2}}\left[  B^{2}+4+\frac{6}{R^{2}}\right]   &  \leq
B\ \ ,\label{ban10.7}\\
\frac{B}{2} +\frac{3}{8R^{2}}\left[  B^{2}+4+\frac{6}{R^{2}}\right]   &  \leq
B\ . \label{ban11.7}%
\end{align}
Therefore, in order for this theorem to work, the length $L$ of the
tube-shaped region in which these multi-Skyrmions are living cannot exceed the
bounds defined in Eqs. (\ref{ban10.5}) and (\ref{ban11.5}) (indeed, if $L$ is
too large, the bounds will be violated at a certain point\footnote{the fact
that $L$ cannot be arbitrarily big for the theorem to hold can be also seen by
observing that in the limit $L\rightarrow\infty$ the domain of definition of
the functions would not be compact any more, thus invalidating the
Ascoli-Arzel\`{a} theorem}.). It is also to be noticed that one cannot obtain
a very large value for the allowed $L$ by increasing $B$ since the left hand
sides of Eqs. (\ref{ban10.5}) and (\ref{ban11.5}) increase faster than the
right hand sides. Moreover, the situation gets worse if $R^{2}$ is very small
(as all the above inequalities are violated at a certain point). However, if
$R$ is large (namely, in the flat limit), Eqs. (\ref{ban10.7}) and
(\ref{ban11.7}) are always satisfied and Eqs. (\ref{ban10.6}) and
(\ref{ban11.6}) become mild constraints on the initial data. On the other
hand, it is trivial to see that it is always possible to choose the initial
data and $B$, $L$ and $R$ in such a way that all the above inequalities are fulfilled.

\textit{The second step} to prove that $T$ is compact is to show that if
$\overrightarrow{F}_{n}\left(  x\right)  =\left(  F_{0}^{n}\left(  x\right)
,F_{1}^{n}\left(  x\right)  \right)  $ is a sequence in $C$ then the sequence
$\overrightarrow{T}\left[  \ \overrightarrow{F}_{n}\left(  x\right)  \right]
$ is \textit{equicontinuous}. To show this, we must evaluate, \textit{for a
generic} $n$, the absolute values of following differences:%
\begin{align}
\left\vert T_{0}\left(  n;x\right)  -T_{0}\left(  n;y\right)  \right\vert  &
=\left\vert b_{0}\left(  x-y\right)  +\int_{x}^{y}H_{0}\left(  n;s\right)
ds\right\vert \ ,\label{ban12}\\
\left\vert T_{1}\left(  n;x\right)  -T_{1}\left(  n;y\right)  \right\vert  &
=\left\vert b_{1}\left(  x-y\right)  +\int_{x}^{y}H_{1}\left(  n;s\right)
ds\right\vert \ , \label{ban13}%
\end{align}
where $0<x<y<L$. After trivial manipulations (which use the fact that all
the functions $\overrightarrow{F}_{n}\left(  x\right)  $ belong to $C$ and
consequently Eqs.(\ref{ban6}) and (\ref{ban7}) are satisfied) one arrives at%

\begin{align}
\left\vert T_{0}\left(  n;x\right)  -T_{0}\left(  n;y\right)  \right\vert  &
\leq\left\vert x-y\right\vert \left[  \left\vert b_{0}\right\vert +L\left\{
\frac{1}{2}B+ \frac{3}{8R^{2}}\left[  B^{2}+4+\frac{6}{R^{2}}\right]
\right\}  \right]  \ ,\label{ban14}\\
\left\vert T_{1}\left(  n;x\right)  -T_{1}\left(  n;y\right)  \right\vert  &
\leq\left\vert x-y\right\vert \left[  \left\vert b_{1}\right\vert +L\left\{
\frac{1}{2}B+ \frac{3}{8R^{2}}\left[  B^{2}+4+\frac{6}{R^{2}}\right]
\right\}  \right]  \ , \label{ban15}%
\end{align}

Thus, given any $\epsilon>0$, one can choose%
\begin{equation}
\delta<\frac{\epsilon}{2\left\{  \left\vert b_{0}\right\vert +\left\vert
b_{1}\right\vert +L\left\{  \frac{1}{2}B+ \frac{3}{8R^{2}}\left[
B^{2}+4+\frac{6}{R^{2}}\right]  \right\}  \right\}  }\ , \label{ban16}%
\end{equation}

in such a way that \textit{both} the choice of $\delta$ in Eq. (\ref{ban16})
does not depend on $n$ \textit{and}, moreover,
\begin{equation}
\left\vert x-y\right\vert <\delta\quad\Rightarrow\quad\left\vert T_{i}\left(
n;x\right)  -T_{i}\left(  n;y\right)  \right\vert \leq\frac{\epsilon}%
{2}\ ,\ \ \forall\ n,\,\, i=0,1\ . \label{ban18}%
\end{equation}

In summary, Eqs. (\ref{ban10}), (\ref{ban11}), (\ref{ban10.5}), (\ref{ban11.5}%
), (\ref{ban10.6}), (\ref{ban11.6}), (\ref{ban10.7}) and (\ref{ban11.7}) show
that, if $\overrightarrow{F}_{n}\left(  x\right)  =\left(  F_{0}^{n}\left(
x\right)  ,F_{1}^{n}\left(  x\right)  \right)  $ is any sequence in $C$, then
the sequence $\overrightarrow{T}\left[  \ \overrightarrow{F}_{n}\left(
x\right)  \right]  $ is uniformly bounded in $C$. Secondly, Eqs.
(\ref{ban14}), (\ref{ban15}), (\ref{ban16}) and (\ref{ban18}) show that, if
$\overrightarrow{F}_{n}\left(  x\right)  =\left(  F_{0}^{n}\left(  x\right)
,F_{1}^{n}\left(  x\right)  \right)  $ is any sequence in $C$, then the
sequence $\overrightarrow{T}\left[  \ \overrightarrow{F}_{n}\left(  x\right)
\right]  $ is \textit{equicontinuous}. Consequently, by virtue of the
Ascoli-Arzel\`{a} theorem, from any sequence $\overrightarrow{T}\left[
\ \overrightarrow{F}_{n}\left(  x\right)  \right]  $ one can extract a
convergent subsequence: this, together with the bounds (\ref{ban10.5}%
)-(\ref{ban11.7}), implies that the operator $\overrightarrow{T}$ is a compact
operator from a bounded closed convex set into itself. Thus, it is possible to
apply the Schauder theorem, which ensures that Eq. (\ref{fixedpoint}) (which
is equivalent to our original system) has at least one solution. Moreover, it
is always possible to choose appropriately the inital data $a_{i}$ and $b_{i}$
in such a way that the two profiles are not proportional.

The conclusion is that not only one can construct numerical solutions (which
are very interesting by themselves), as we do in the main text, but also one
can prove analytically that non-trivial multi-Skyrmions in which the profiles
are not proportional do indeed exist. Besides the intrinsic mathematical
elegance of the fixed-point Schauder-type argument, the present procedure also
discloses the presence of the bounds in Eqs.(\ref{ban10.5})-(\ref{ban11.7}) on
the radius $R$ and on the length $L$ of the tube-shaped region in which these
multi-Skyrmions are living (which constrain its shape), as well as on the other parameters of the model.
At the present stage of the analysis, it is not possible yet to say whether
such a bound is just a limitation of the method or it signals some deeper
physical limitation on the volume of the regions in which one constrains these
Skyrmions to live. However, to understand whether or not $SU(N)$
multi-Skyrmions can fit into very large tube-shaped regions is certainly a
very interesting and deep question on which we hope to come back in a future investigation.

\end{document}